\documentclass[11pt,a4paper]{article}
\usepackage{amsmath,amssymb}
\usepackage[pdftex]{graphicx}
\usepackage{float}
\usepackage{fancybox,arydshln,multirow,bigdelim}
\usepackage{color}

\makeatletter
\renewcommand{\section}{%
 \@startsection{section}{1}{\z@}{-3.5ex \@plus -1ex \@minus -.2ex}{2.3ex \@plus.2ex}%
  {\normalfont\Large\bfseries}}
\makeatother

\makeatletter
\renewcommand{\subsection}{%
 \@startsection{subsection}{1}{\z@}{-3.5ex \@plus -1ex \@minus -.2ex}{2.3ex \@plus.2ex}%
  {\normalfont\large\bfseries}}
\makeatother

\makeatletter
\def\appendix{
\def\theequation{\Alph{section}\arabic{equation}}
\setcounter{equation}{0}
\def\thesection{\Alph{section}}
\@addtoreset{equation}{section}
\setcounter{section}{0}
\def\@seccntformat##1{
\@nameuse{prefix@##1}
\@nameuse{the##1}
\@nameuse{postfix@##1}\quad}
\def\prefix@section{Appendix~}
} 
\makeatother

\def\slashchar#1{\setbox0=\hbox{$#1$}
\dimen0=\wd0 
\setbox1=\hbox{/} \dimen1=\wd1 
\ifdim\dimen0>\dimen1 
\rlap{\hbox to \dimen0{\hfil/\hfil}} 
#1 
\else 
\rlap{\hbox to \dimen1{\hfil$#1$\hfil}} 
/ 
\fi}

\allowdisplaybreaks[2] 

\def\bx{{\boldsymbol x}}

\def\bq{{\boldsymbol q}}
\def\bp{{\boldsymbol p}}
\def\bE{{\boldsymbol E}}
\def\bB{{\boldsymbol B}}

\def\cd{\cdot}
\newcommand{\ltsim}{\protect\raisebox{-0.5ex}{$\:\stackrel{\textstyle <}{\sim}\:$}}
\newcommand{\gtsim}{\protect\raisebox{-0.5ex}{$\:\stackrel{\textstyle >}{\sim}\:$}}

\begin{document}

\title{{\bf Electromagnetic currents induced by color fields} }
\author{Naoto Tanji\footnote{\textit{E-mail address}: tanji@thphys.uni-heidelberg.de }}
\date{{\it Institut f{\"u}r Theoretische Physik, Universit{\"a}t Heidelberg \\
Philosophenweg 16, 69120 Heidelberg, Germany}}

\maketitle

\begin{abstract}
The quark production in classical color fields is investigated with a focus on 
the induction of an electromagnetic current by produced quarks.  
We show that the color SU(2) and the SU(3) theories lead significantly different results 
for the electromagnetic current. 
In uniform SU(2) color fields, the net electromagnetic current is not generated,
while in SU(3) color fields the net current is induced depending on the color direction of background fields.  
Also the numerical study of the quark production in inhomogeneous color fields is done.
Motivated by gauge field configurations provided by the color glass condensate framework, 
we introduce an ensemble of randomly distributed color electric fluxtubes.  
The spectrum of photons emitted from the quarks by a classical process is shown. 
\end{abstract}

\section{Introduction} \label{sec:intro}

Photons  are one of the most important observables 
to explore the properties of high-energy matter produced in 
relativistic heavy-ion collisions, 
since they are penetrating probes which are free from strong interactions. 
In recent years, direct photon measurement by the PHENIX \cite{Adare:2008ab,Adare:2011zr}
and the ALICE Collaborations \cite{Wilde:2012wc,Lohner:2012ct} posed a puzzle 
to theoretical understanding of photon production.
The direct photon puzzle consists of two issues. 
(i) State-of-the-art theoretical calculations 
\cite{vanHees:2011vb,Linnyk:2013wma,Shen:2013vja,Shen:2013cca} underestimate 
direct photon excess measured in experiments \cite{Adare:2008ab,Wilde:2012wc}. 
(ii) A large elliptic flow of direct photons comparable in magnitude to that of pions are observed \cite{Adare:2011zr,Lohner:2012ct}. 
It is difficult to explain these two issues simultaneously
because a large direct photon yield favors photon emission at early time 
when temperature of the system is high,
while a large elliptic flow indicates that photons are emitted at later time
when the flow of matter is largely developed. 
This puzzle naturally drives people to think about a novel mechanism of photon production,
e.g. \cite{Chiu:2012ij,McLerran:2014hza,Basar:2012bp,Tuchin:2014pka,Gale:2014dfa}. 

In Ref.~\cite{Klein-Bosing:2014uaa}, it has been shown that
the photon spectra measured in p$+$p, d$+$Au, Au$+$Au collisions at RHIC 
and Pb$+$Pb collisions at LHC satisfy geometric scaling.
This result indicates that the photons are emitted in early time
when the saturation momentum $Q_s$ is the only dominant scale of the system.
In the leading order description of the color glass condensate, 
the system right after a heavy-ion collision consists of
longitudinal color electric and magnetic fields \cite{Kovner:1995ja,Kovner:1995ts,Lappi:2006fp}. 
This state is called Glasma. 
The photon production in the Glasma has been studied in Refs.~\cite{Chiu:2012ij,McLerran:2014hza}. 
Although promising results are presented,
these studies are based on a simplified model for the gluon distribution function in the Glasma. 
To obtain deeper understanding of the photon production in the Glasma, 
calculations based on a first principle is needed.  
The purpose of the present study is to take a first step toward this direction. 

In order to investigate the photon production in the Glasma 
(the central rapidity region of nucleus-nucleus collisions), 
we need to first compute quark production because the initial states described by the color glass condensate 
(see e.g. \cite{Iancu:2002xk,Weigert:2005us,Gelis:2010nm}) purely consist of gluons, 
which do not directly couple to photons.
This situation is in contrast to proton-nucleus collisions, where the proton can be viewed 
as high-energy partons, not as classical gauge fields.   
The photon and dilepton production in proton-nucleus collisions have been investigated 
by calculating the processes that a quark in the proton emits a photon due to the interaction with 
strong color classical fields of the nucleus \cite{Gelis:2002ki,Gelis:2002fw,Baier:2004tj,Kovner:2014qea}.

In the leading order description of the color glass condensate and the Glasma, 
gauge fields are treated as classical fields
because the coupling is small, $g\ll 1$, and the fields are strong, $A\sim 1/g$ (see e.g. \cite{Gelis:2012ri}). 
The quark dynamics in the leading order is described by retarded solutions of
the Dirac equation under the classical gauge field. 
For generic inhomogeneous gauge fields, we need to resort to lattice-discretized numerical calculations
to solve the Dirac equation. 
In studies of real-time dynamics of fermions on lattice, 
there have been considerable advances 
\cite{Aarts:1998td,Borsanyi:2008eu,Berges:2013oba,Kasper:2014uaa,Gelis:2015eua}. 
Given these advances, we will give a first-principle description of the quark dynamics
under classical gauge fields, 
and investigate the induction of electromagnetic (EM) current 
accompanying the quark production. 
In this paper, we assume a system in a fixed box, namely 
boost-invariant expansion is not taken into account. 
Besides we focus on only the one-point expectation value of the EM current operator.
From the expectation of the current, the spectrum of photons
produced by a classical process can be computed.

In the context of real-time lattice simulations of the early-time dynamics 
in heavy-ion collisions, 
the color SU(2) theory is often used
instead of the color SU(3) just for simplicity,
and it is sometimes expected that SU(3)${}_c$ results would be similar to SU(2)${}_c$. 
Indeed,  it has shown that the dynamics of classical gauge fields exhibits no qualitative difference 
between SU(2)${}_c$ and SU(3)${}_c$ \cite{Berges:2008zt}. 
We will show that, however, 
this is not the case for the induction of EM currents.
In SU(2)${}_c$, there is a tendency that two color components contribute to the EM current 
with opposite signs to each other. 
We will show that this cancellation is exact in uniform fields. 
On the other hand, in SU(3)${}_c$, the degree of the cancellation between different color components
depends on the color direction of background fields.
Thus, the EM current can be several orders larger than that in SU(2)${}_c$. 

The rest of this paper is organized as follows.
The induction of EM currents by uniform color electric fields is discussed in Sec.~\ref{sec:uniform}.
We compare lattice numerical results with known analytic solutions
to show the validity of our numerical method.
In Sec.~\ref{sec:glasma}, we investigate the quark production and the subsequent induction of EM currents 
in inhomogeneous background fields, whose configuration mimics the Glasma.
The photon spectrum associated with a classical emission process is shown at the last.
Section \ref{sec:summary} is devoted to summary and discussions. 
The lattice numerical method we employ is explained in the Appendix. 
In particular, our method to treat doubler modes is described in detail.

\section{Uniform color electric fields} \label{sec:uniform}
In this section, we investigate the induction of EM currents in uniform color electric fields. 
Such a simple configuration for color fields enables us to do analytical studies on 
real-time dynamics of quark production 
via the Schwinger mechanism \cite{Tanji:2008ku,Tanji:2010eu},
and thus clarifies the difference between SU(2) and SU(3) color fields for the induction of EM currents.
Therefore, studying the uniform color field configurations can provide us with useful information
to understand more complicated and realistic situation,
which is discussed in the next section.  

Furthermore, the uniform field configurations can be used as a benchmark test for 
the real-time lattice numerical computations. 
By comparing numerical results with known analytic results, we can check the validity
of our numerical method on lattice before applying it to field configurations 
that analytical solutions are not available. 
See Appendix \ref{sec:numerical} for details of the lattice numerical method. 

In this paper, we consider only one family of quark flavor which has an EM charge $e$ and mass $m$,
for simplicity.
We will briefly comment on the effects of multi-flavors 
in Sec.~\ref{sec:summary}. 

\subsection{Abelianization} \label{subsec:Abel}
We briefly review a method to simplify the color structure of the Dirac equation
under uniform color electromagnetic fields \cite{Tanji:2010eu,Nayak:2005pf}. 
Thanks to this simplification, the induction of EM currents in uniform color fields can be
understood intuitively. 

If the field strength is spatially uniform, its color dependence can be factorized as
\begin{equation}
F_{\mu\nu}^a = F_{\mu\nu} n^a \, ,
\end{equation}
where $n^a$ is a constant vector in color space. 
This kind of fields are called covariant constant fields \cite{Batalin:1976uv,Gyulassy:1986jq}. 
This field strength is given by the gauge field
\begin{equation} \label{gaugefield}
A_\mu^a (x) = A_\mu (x) n^a \, ,
\end{equation}
where $A_\mu (x)$ is related with $F_{\mu \nu}$ by $F_{\mu\nu} = \partial_\mu A_\nu -\partial_\nu A_\mu$. 
The gauge field \eqref{gaugefield} is a solution of the Yang--Mills equation. 
Because the color vector $n^a$ is constant, the structure of 
the covariant derivative $D_\mu = \partial_\mu -igA_\mu (x) n^a T^a$ can be reduced
from SU($N_c$) to $[$U(1)$]^{N_c}$. 
Since $n^a T^a$ is an $N_c \times N_c$ hermitian matrix, one can diagonalize it by a unitary transformation as
\begin{equation}
Un^a T^a U^\dagger = \text{diag} \left( w_1 ,\cdots ,w_{N_c} \right) . 
\end{equation}
The color-rotated quark field $\psi_i ^\prime = \left( U\psi \right)_i $ $(i=1,\cdots ,N_c )$ obeys
\begin{equation}
\left[ i\gamma^\mu \left( \partial_\mu -iw_i gA_\mu \right) -m \right] \psi_i ^\prime = 0 \, ,
\end{equation} 
which has the same structure as the U(1) theory but 
a coupling $e$ is replaced by the effective coupling $w_i g$. 

The SU(2) color space is isotropic since the rank of SU(2) is one. 
For any unit color vector $n^a$, 
the color matrix $n^a T^a$ takes the same form after diagonalization:
\begin{equation}
Un^a T^a U^\dagger = \text{diag} \left( \frac{1}{2} ,-\frac{1}{2} \right) . 
\end{equation}
Therefore, the effective couplings are always $+g/2$ and $-g/2$. 

For $N_c \geq 3$, the color space is anisotropic
in a sense that the effective couplings depend on the direction of the color vector $n^a$ in a gauge-invariant way. 
For SU(3)${}_c$, there are two independent diagonal generators. 
If we take the generators $T^a =\frac{1}{2} \lambda^a$ with the Gell-Mann matrix $\lambda^a$,
the diagonal generators are $T^3$ and $T^8$, 
by which the diagonalized color matrix $n^a T^a$ can be expanded:
\begin{equation}
Un^a T^a U^\dagger = T^3 \cos \theta -T^8 \sin \theta . 
\end{equation}
The angle $\theta$ is related with the color vector $n^a$ through the second Casimir invariant 
$C_2 =\left[ d^{abc} n^a n^b n^c \right]^2$ as \cite{Nayak:2005pf}
\begin{equation}
\sin^2 3\theta = 3C_2 \, . 
\end{equation}
By this equation, the angle $\theta$ characterizes the color direction of the vector $n^a$
in a gauge-invariant way. 
The eigenvalues of the matrix $n^a T^a$ are parametrized by $\theta$ as
\begin{align} 
w_1 &=\frac{1}{\sqrt{3}} \cos \left( \theta +\frac{\pi}{6} \right) \, , \notag \\
w_2 &=\frac{1}{\sqrt{3}} \cos \left( \theta +\frac{5\pi}{6} \right) \, , \label{effective} \\
w_3 &=\frac{1}{\sqrt{3}} \cos \left( \theta +\frac{3\pi}{2} \right) \, . \notag
\end{align} 
Unlike the SU(2)${}_c$ case, these effective couplings have the color direction dependence.

\subsection{EM currents in uniform color electric fields} \label{subsec:EMc_uni}

\begin{figure}[bp]
\begin{center}
\includegraphics[clip,width=10cm]{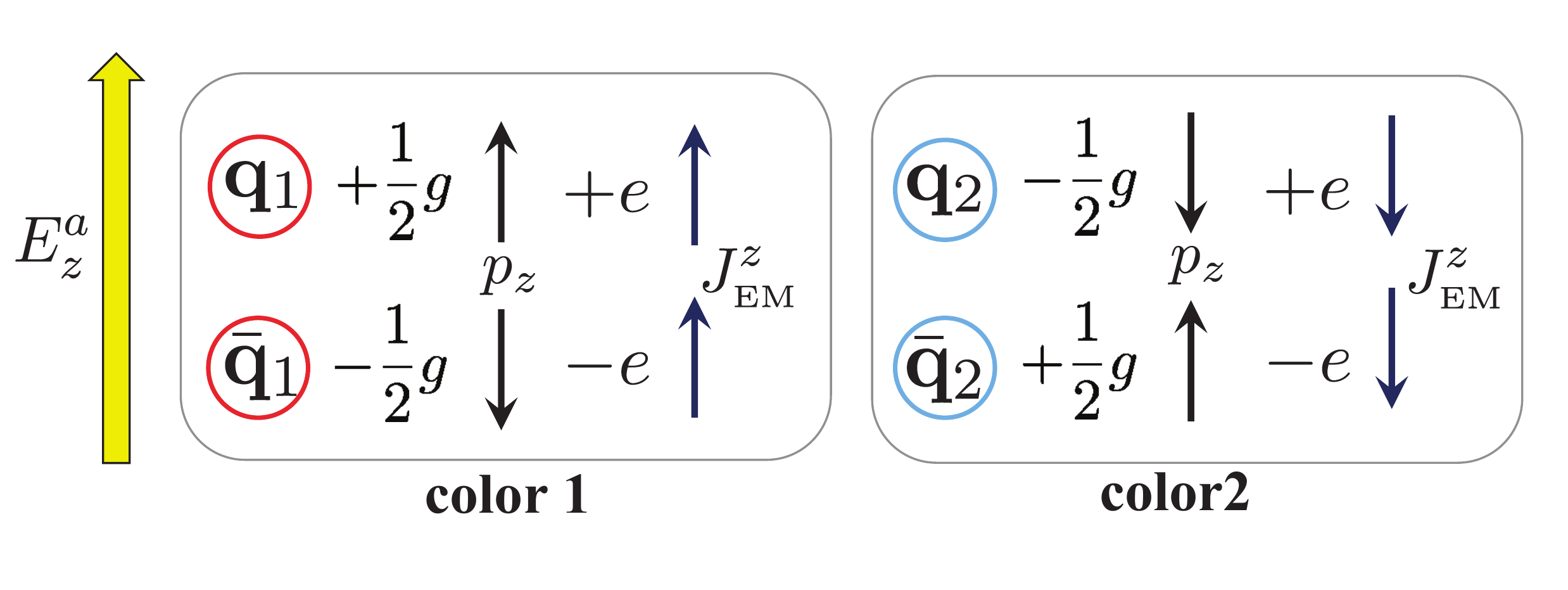}
\caption{SU(2)${}_c$ quarks in a constant color electric field.  }
\label{fig:su2}
\end{center}
\end{figure}

Having discussed the diagonalization (Abelianization) of the color matrix, 
one can easily prove the cancellation of EM currents in SU(2)${}_c$ uniform electric fields
(see Fig.~\ref{fig:su2}). 
Let us suppose that the electric field is applied to the $z$-direction with the strength $E_0$ $(>0)$. 
In the case of SU(2)${}_c$, 
the effective couplings are independent of the color direction of the gauge field.
A quark with color 1 couples to the electric field with the effective coupling of $+\frac{1}{2} g$,
and it also has an EM charge $+e$ (we assume $e>0$). 
This quark is accelerated to the $+z$-direction by the electric field
and causes a positive EM current. 
An anti-quark is accelerated to the opposite direction but it has a negative EM charge,
so that it also induces a positive EM current. 
On the other hand, a quark with color 2 has the effective coupling $-\frac{1}{2}g$.
Therefore, this quark is accelerated to $-z$-direction. 
Although the color charges are opposite in sign between color 1 and 2, 
the EM charge is independent of the color components.
Consequently, the quark with color 2 induces a negative current, 
which cancels the current induced by the quark with color 1. 
The same cancellation happens for anti-quarks. 
In this way, we can conclude that uniform SU(2) color electric fields
do not induce an EM current. 

Clearly, this argument does not apply to the SU(3)${}_c$ case.
Although there is alway a tendency of the cancellation among
different color components, 
the cancellation may not be perfect. 
When $\theta =0$ (modulo $\pi/3$), SU(3)${}_c$ is reduced to SU(2)${}_c$
because one of three quarks decouples from the color field.
For other values of the color angle $\theta$, 
a nonzero EM current is induced as we will show explicitly 
in the next subsection.

\subsection{Numerical results} \label{subsec:results_uni}
In the following we consider a uniform color electric field turned on at $t=0$:
\begin{equation} \label{constE}
F_{03}^a = \begin{cases} E_0 n^a & (t\geq 0 ) \\ 0 & (t<0). \end{cases}
\end{equation}
This field configuration is a solution of the Yang--Mills equation coupled to a pulse-like external current
\begin{equation}
J_\text{ext}^{a\, i} = -E_0 n^a \delta^{i3} \delta (t) \, . 
\end{equation}

\begin{figure}[tbp]
\begin{center}
\includegraphics[clip,width=8.0cm]{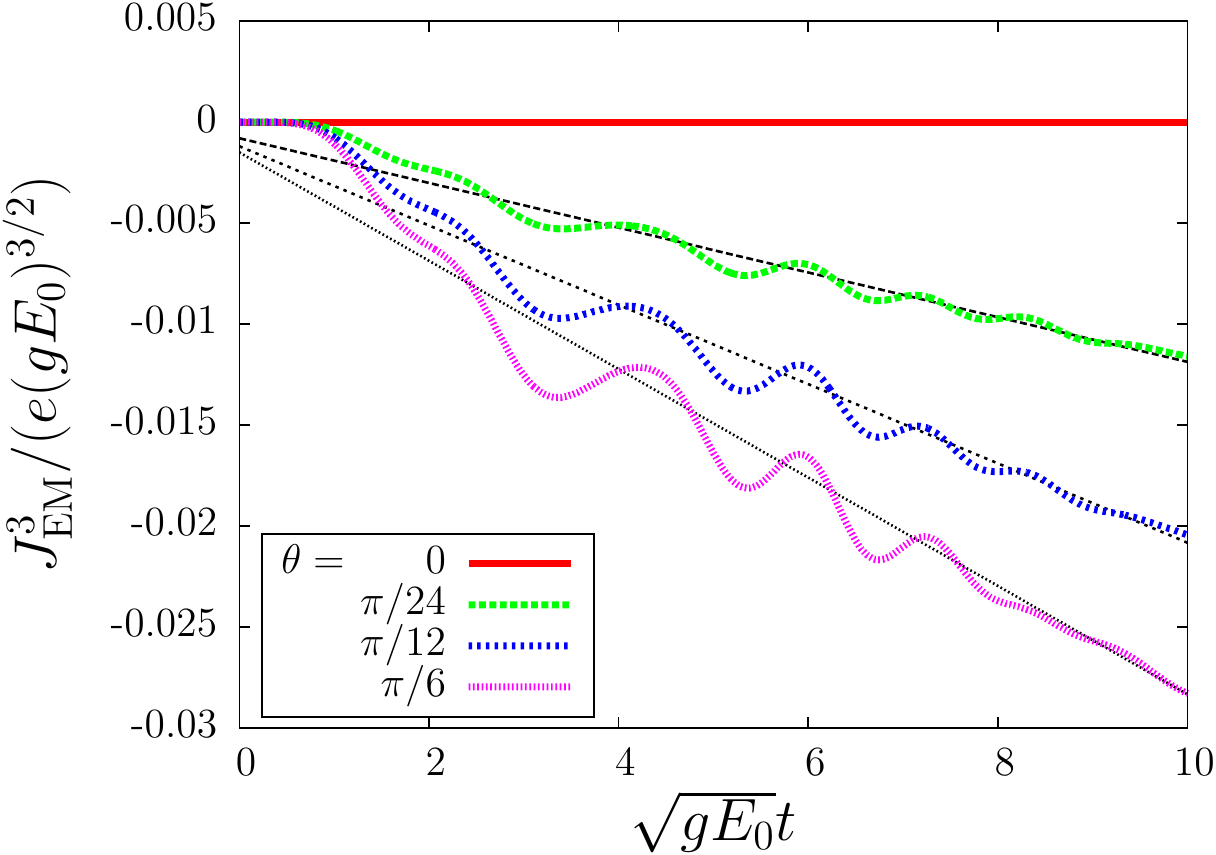}
\caption{The time dependence of the EM current induced by the uniform electric field for various values of the color angle $\theta$. 
 The lattice numerical results (bold lines) are compared with the simple estimation \eqref{EMc_model} (thin black lines). 
 Parameters used for these numerical computations are 
 $m/Q_s=0.1$, $Q_s L_x =Q_s L_y =30$, $Q_s L_z=15$, $N_x=N_y=48$, and $N_z=192$.}
\label{fig:EMc_uni}
\end{center}
\end{figure}
In Fig.~\ref{fig:EMc_uni}, the expectation value of 
the EM currents obtained by lattice numerical calculations are plotted
for some different values of the color angle $\theta$ in the case of SU(3)${}_c$.
Only the $z$-component of the current is shown because
the $x$- and $y$-components are zero by the symmetry of the system. 
When $\theta =0$, the EM current is indeed zero 
because SU(3)${}_c$ is reduced to SU(2)${}_c$. 
For other values of $\theta$, a nonzero EM current is generated 
and its magnitude largely depends on the value of $\theta$. 
This strong dependence of the EM current on the color angle $\theta$
stands in contrast to the weak dependence on $\theta$ that
the the total quark number density shows \cite{Cooper:2008af,Tanji:2010eu}. 
This result indicates that EM observables may be more sensitive to color field configurations
than hadronic observables. 

The linear increase of the EM current seen in Fig.~\ref{fig:EMc_uni} and its dependence on $\theta$ 
can be understood from analytical results. 
Thanks to the Abelianization, the Dirac equation under the uniform electric field \eqref{constE} 
can be analytically solved, and the distribution function of produced quarks
is available in a closed form \cite{Tanji:2010eu}. 
Although the distribution function has a complicated expression in terms of special functions, 
its behavior is well described by the following simple function \cite{Tanji:2008ku}:
\begin{equation} \label{model}
f_{s,c} (t;\bp ) = \exp \left( -\pi \frac{m^2 +p_x^2 +p_y^2}{|w_c g|E_0 } \right)
 \Theta (p_z ;0,w_c gE_0 t ) \, , 
\end{equation} 
where the function $\Theta (x;a,b)$ is defined by
\begin{equation}
\Theta (x;a,b) = \begin{cases} 1 & \text{min}(a,b) \leq x \leq \text{max}(a,b) \\
                               0 & \text{otherwise}.
                 \end{cases} 
\end{equation} 
The expectation value of the EM current can be expressed 
by the analytic expressions for the mode functions. 
However, it still contains the integration over momentum, 
which needs to be evaluated numerically,
making the parametric dependence on the color angle hardly accessible.
Instead of doing the numerical integration, we use the simple expression \eqref{model}
to evaluate the EM current, and compare it with the results of lattice numerical calculations. 
In terms of the distribution function, the EM current can be expressed as
\begin{equation} \label{Jcond}
J_\text{{\tiny EM}}^i (t) 
 = 2e\sum_{s,c} \int \frac{d^3 p}{(2\pi )^3} \frac{p^i}{\omega_p } f_{s,c} (t;\bp ) \, . 
\end{equation}
This is called conduction current, 
which is caused by movement of charged particles. 
In addition to the conduction current, there is another contribution to the current, 
called polarization current, 
which cannot be expressed by the normal distribution function. 
Except the times right after the turning-on of the electric field, the polarization current
is negligible compared with the conduction current under a strong field, 
$gE \gg m^2$ \cite{Tanji:2008ku}.
We therefore consider only the conduction current. 
If we substitute Eq.~\eqref{model} into Eq.~\eqref{Jcond}, 
and assume $p_z /\omega_p \simeq \text{sgn} (p_z )$, which is valid for $|w_c g|E_0 t \gg m$, 
we get a simple expression for the longitudinal EM current,
\begin{equation} \label{EMc_model} 
J_\text{{\tiny EM}}^3 (t) 
 \simeq \frac{4e}{(2\pi)^3} \sum_c e^{-\pi \frac{m^2}{|w_c g|E_0}} w_c g |w_c g| E_0^2 t \, . 
\end{equation}
This simple estimation is plotted in Fig.~\ref{fig:EMc_uni} as thin black lines 
with nonzero constants added to fit the numerical data. 
Eq.~\eqref{EMc_model} well reproduces the $\theta$-dependence that
the numerical results show, especially at later times. 

\begin{figure}[tbp]
\begin{center}
\includegraphics[clip,width=8.0cm]{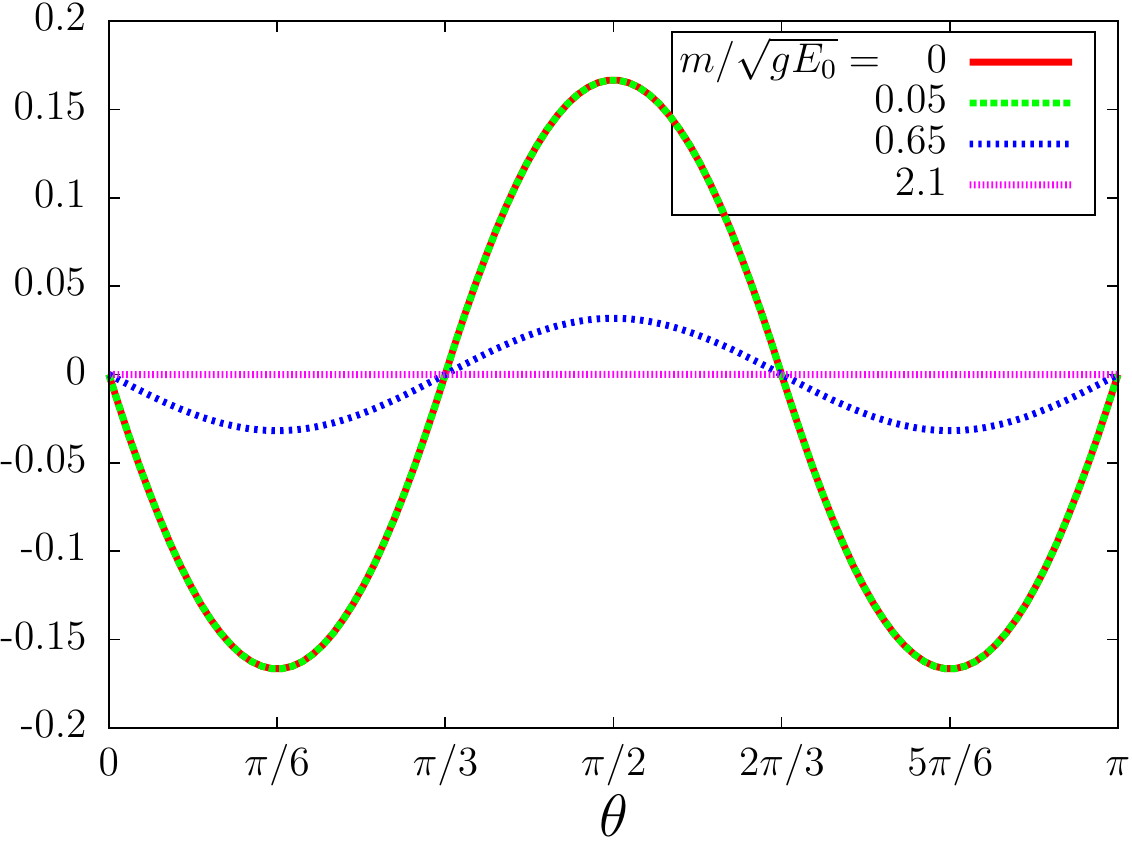}
\caption{The color angle $\theta$ dependence of 
the coefficient of the linear increase of the current \eqref{EMc_model},
$\sum_c e^{-\pi \frac{m^2}{|w_c g|E_0}} w_c g |w_c g|$, for various quark mass.}
\label{fig:coefficient}
\end{center}
\end{figure}
In Fig.~\ref{fig:coefficient}, the coefficient of the linear increase of the current \eqref{EMc_model}
is plotted as a function of the color angle $\theta$ for various quark mass.
As indicated by Eqs.~\eqref{effective}, this is a periodic function with a period of $2\pi/3$. 

Before closing this section, let us emphasize the fact that 
the knowledge of the Abelianization has not been used in the numerical computations. 
We have just solved Eqs.~\eqref{eq_U}, \eqref{eq_E} and \eqref{Dirac2}. 
Therefore, this numerical calculation can be easily extended to more complicated gauge field configurations
as we will discuss in the next section. 

\section{Glasma-like electric fields} \label{sec:glasma}
In the previous section, we have dealt with the uniform electric fields
to clarify the physics of the quark production in classical non-Abelian gauge fields
and the difference between SU(2) and SU(3) color fields for the induction of the EM current. 
In this section, we study EM currents induced by 
more realistic configurations for the background gauge fields.

The gauge fields in the Glasma are uniform in the longitudinal direction reflecting
the boost invariance of the system, and are inhomogeneous in the transverse plane \cite{Lappi:2006fp}. 
The inhomogeneity in the transverse plane is characterized by the saturation scale $Q_s$. 
One of the characteristic features of the Glasma is the existence of the longitudinal magnetic fields.
However, the existence of magnetic fields at the initial time is 
inconsistent with the assumption of the free initial condition for spinors. 
In contrast, initial inhomogeneous electric fields can be given by zero gauge field $A(0)=0$, 
and thus they can be consistent with the free initial condition for spinors. 
In the $\tau$-$\eta$ coordinates, one can construct the initial condition for the spinors at $\tau =0$
that is consistent with the evolution of the gauge fields describing a collision of color glass condensates 
\cite{Gelis:2015eua}. 
In this case, the existence of the initial magnetic fields does not cause the inconsistency. 
However, we cannot employ these initial conditions since we assume a system in a non-expanding box. 
Instead of it, we employ the gauge fields that initially consist of only electric fields.   
To mimic the Glasma, we introduce an ensemble of electric fluxtubes. 
In the transverse $x$-$y$ plane, we randomly distribute $N_\text{tube} =Q_s^2 L_x L_y$-numbers of fluxtubes
which has a Gaussian profile with mean radius $1/Q_s$ and which are uniform in the longitudinal direction: 
\begin{equation} \label{randE} 
E_z^a (t=0,\bx_\perp ) = \frac{Q_s^2}{g} \sum_{j=1}^{N_\text{tube}} 
 \exp \left( -Q_s^2 (\bx_\perp -\boldsymbol{\xi}_j )^2 \right) n^a_j \, , 
\end{equation}
where $\boldsymbol{\xi}_j$ is transverse coordinates of a center of a fluxtube, which is randomly distributed in the transverse plane,
and $n^a_j$ is a unit color vector whose orientation is random. 
The distance $|\bx -\boldsymbol{\xi}_j|$ is measured in a way respecting the periodic boundary condition. 
The electric field \eqref{randE} trivially satisfies the Gauss's law \eqref{Gauss}. 

\begin{figure*}[tbp]
 \begin{tabular}{cc}
 \begin{minipage}{0.5\hsize}
  \begin{center}
   \includegraphics[width=7.0cm]{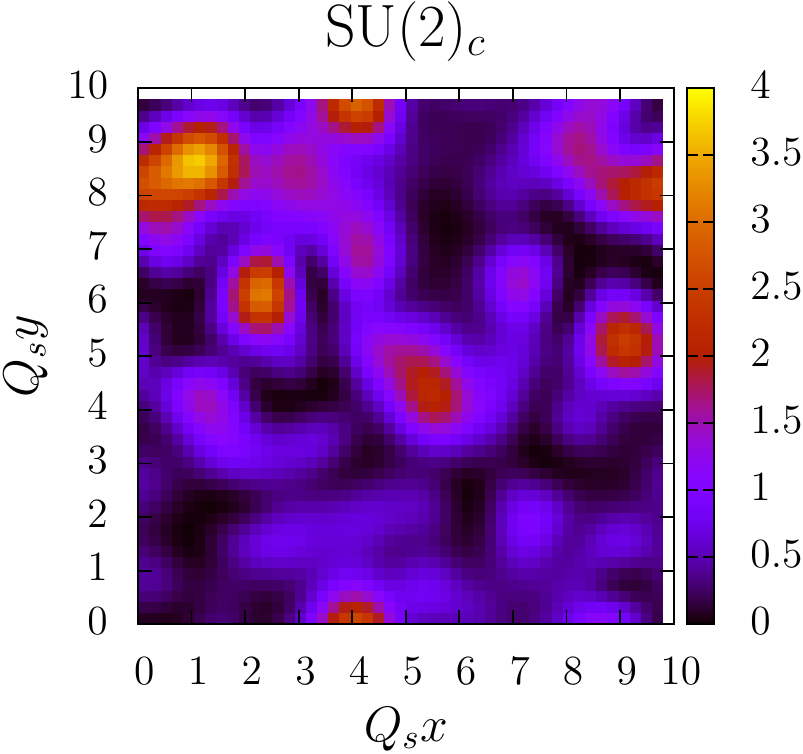} 
  \end{center}
 \end{minipage}
 \begin{minipage}{0.5\hsize}
  \begin{center}
   \includegraphics[width=7.0cm]{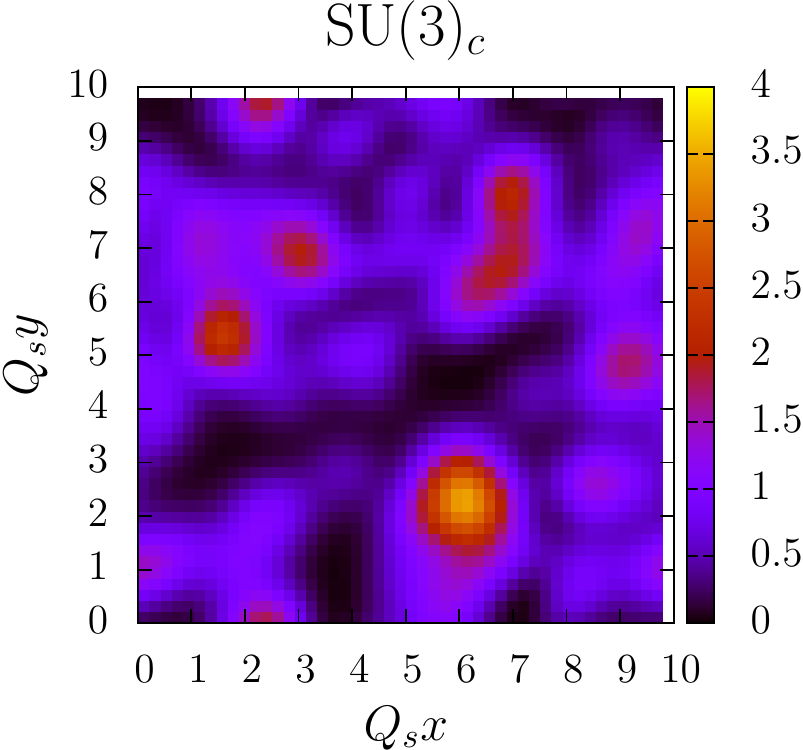} 
  \end{center}
 \end{minipage} 
\end{tabular}
 \caption{\label{fig:gauge_ene2d} 
 Transverse profiles of the initial energy density (times $g^2$) of a randomly generated electric field, 
 $\frac{1}{2} g^2 \bE^2 /Q_s^4$. (Density plots with color bars in the right side of each figure.)
 Left: SU(2)${}_c$, Right: SU(3)${}_c$. 
 Lattice parameters are $Q_s L_x =Q_s L_y =10$, and $N_x=N_y=48$.}
\end{figure*}

\begin{figure*}[tbp]
 \begin{tabular}{cc}
 \begin{minipage}{0.5\hsize}
  \begin{center}
   \includegraphics[width=7.0cm]{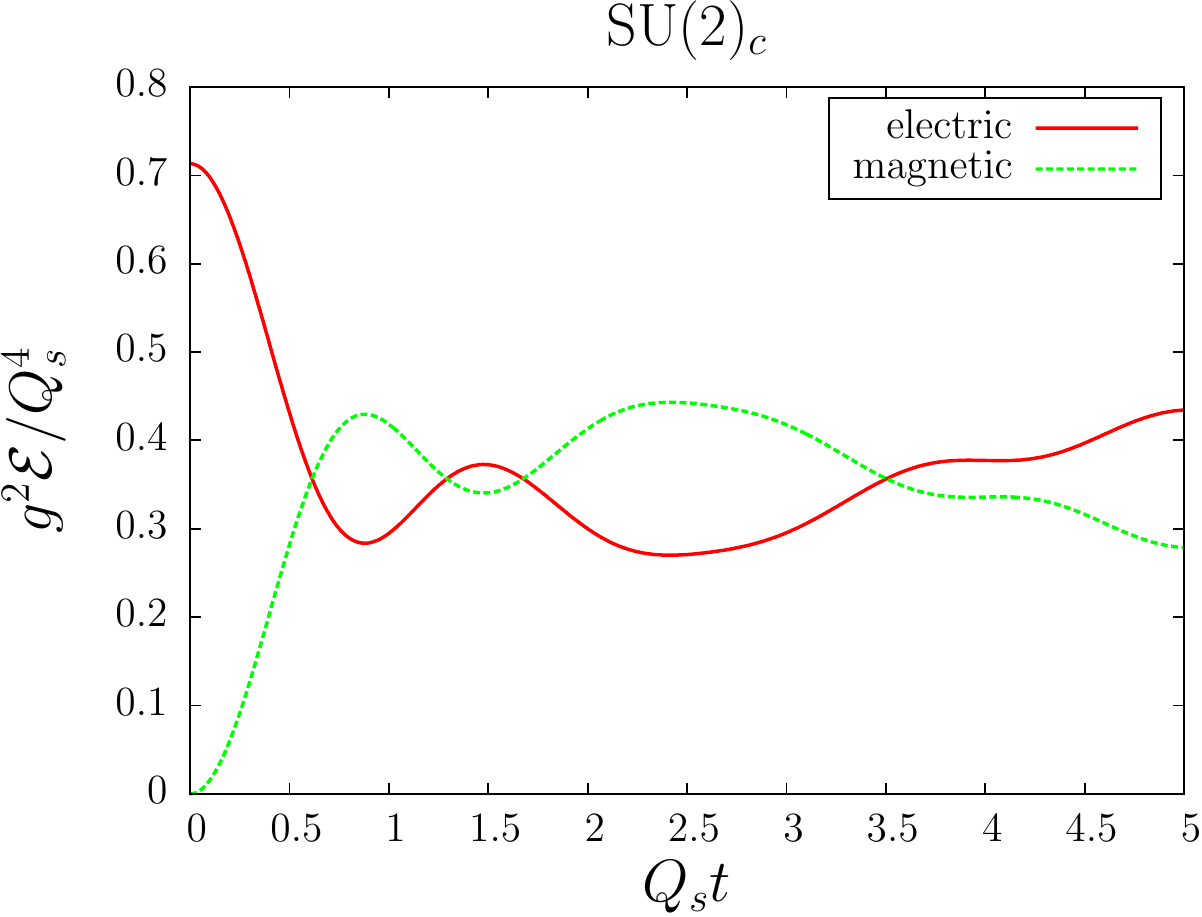} 
  \end{center}
 \end{minipage}
 \begin{minipage}{0.5\hsize}
  \begin{center}
   \includegraphics[width=7.0cm]{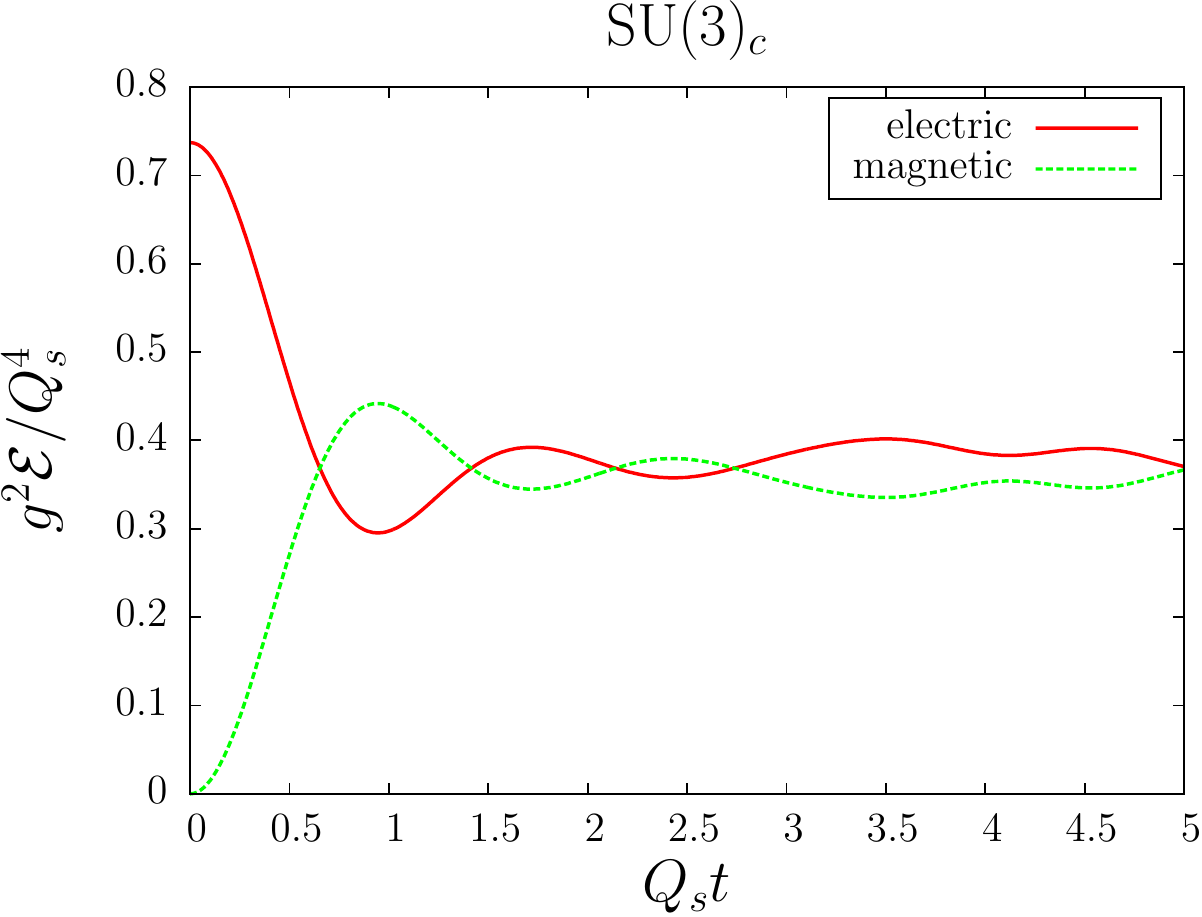} 
  \end{center}
 \end{minipage} 
\end{tabular}
 \caption{\label{fig:gauge_ene1d} Time dependence of the electric and magnetic components of 
 the gauge field energy density averaged over space, 
 $\frac{1}{N^2} \sum_{\bx_T} \frac{1}{2} g^2 \bE^2 /Q_s^4$ 
 and $\frac{1}{N^2} \sum_{\bx_T} \frac{1}{2} g^2 \bB^2 /Q_s^4$. 
 Left: SU(2)${}_c$, Right: SU(3)${}_c$. 
 Lattice parameters are the same as Fig.~\ref{fig:gauge_ene2d}. }
\end{figure*}

Figure \ref{fig:gauge_ene2d} displays the transverse profiles of the initial energy density of
a randomly generated electric field \eqref{randE} for SU(2)${}_c$ and SU(3)${}_c$, respectively. 
Energy density times $g^2$ is plotted since this combination is independent of the value of the coupling $g$. 
In the following, all dimensionful quantities are shown in a dimensionless combination 
scaled by powers of $Q_s$. 
In Fig.~\ref{fig:gauge_ene1d}, the time evolution of the gauge field energy density averaged over volume
is exhibited. 
Owing to the inhomogeneity in the transverse plane, magnetic fields are quickly developed,
and the equipartition of energy between electric and magnetic components is realized
both for SU(2)${}_c$ and SU(3)${}_c$ case.
As observed in Ref.~\cite{Berges:2008zt}, 
the SU(2)${}_c$ and the SU(3)${}_c$ theories show qualitatively similar behavior
for the non-Abelian gauge fields.

\begin{figure*}[tbp]
 \begin{tabular}{cc}
 \begin{minipage}{0.5\hsize}
  \begin{center}
   \includegraphics[width=6.5cm]{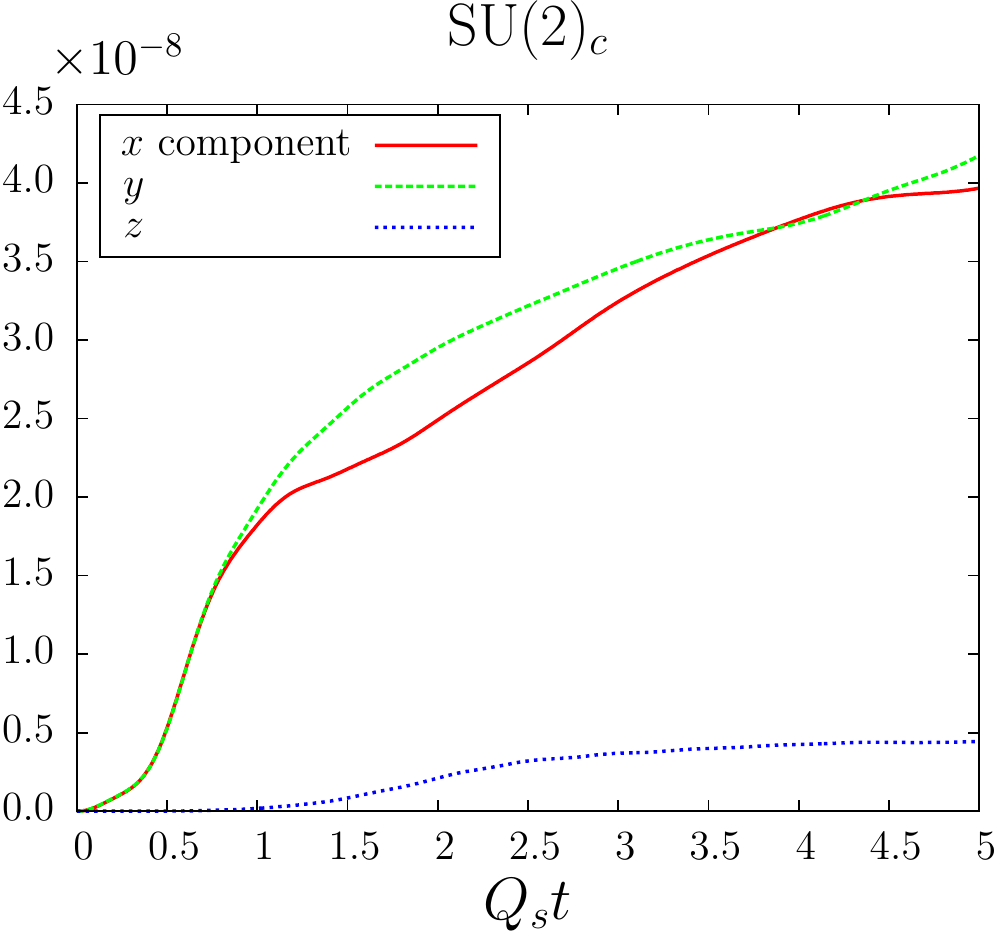} 
  \end{center}
 \end{minipage}
 \begin{minipage}{0.5\hsize}
  \begin{center}
   \includegraphics[width=6.5cm]{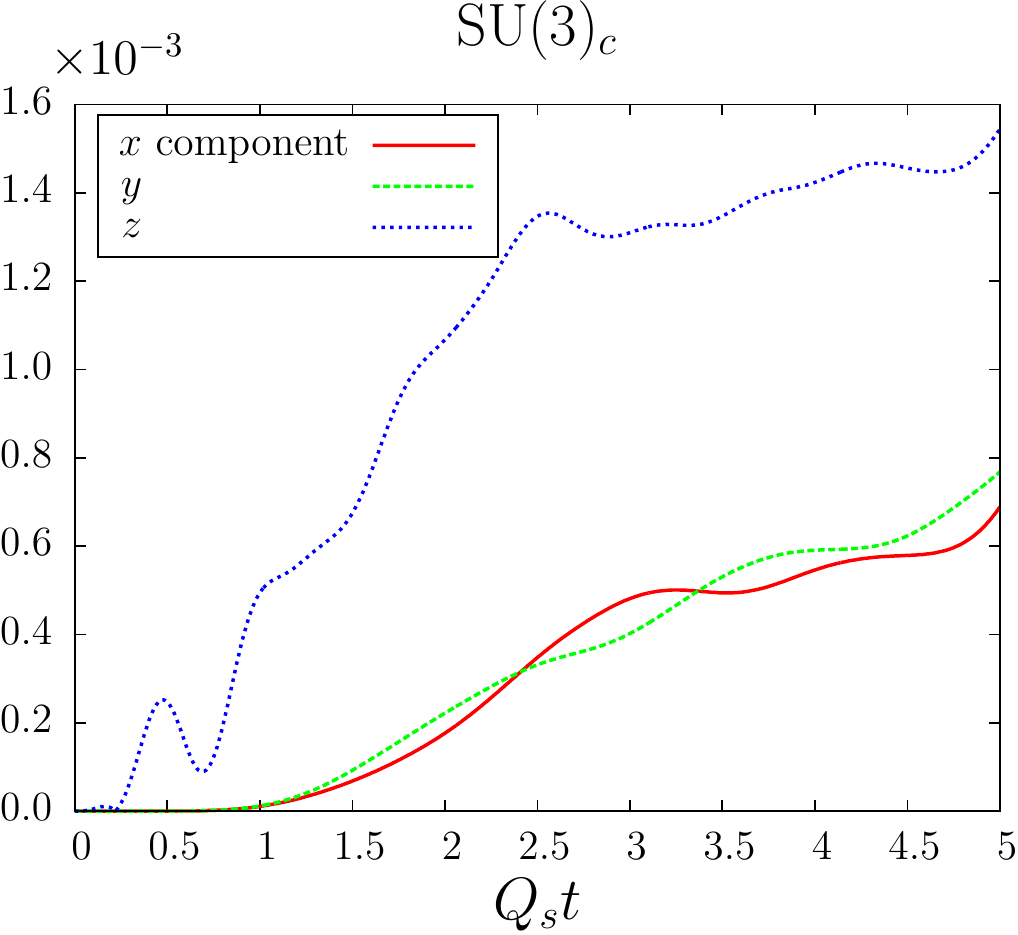} 
  \end{center}
 \end{minipage} 
 \end{tabular}
 \caption{\label{fig:emcsq} 
 Plots of $\sqrt{\frac{1}{N^2} \sum_{\bx_T} \left( J_\text{{\tiny EM}}^i \right)^2}/(e Q_s^3)$ as a function of time.
 Left: SU(2)${}_c$, Right: SU(3)${}_c$. 
 Parameters used for these computations are $m/Q_s=0.1$, $Q_s L_x =Q_s L_y =10$, $Q_s L_z=15$, and
 $N_x=N_y=N_z=48$.  }
\end{figure*}

\begin{figure*}[tbp]
 \begin{tabular}{cc}
 \begin{minipage}{0.5\hsize}
  \begin{center}
   \includegraphics[width=7.0cm]{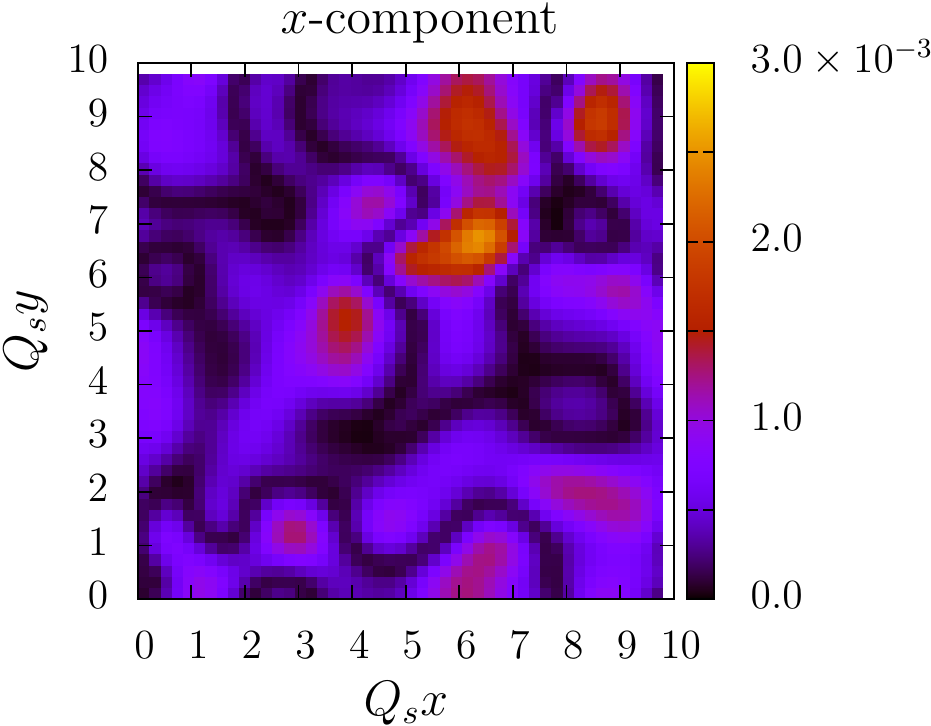} 
  \end{center}
 \end{minipage}
 \begin{minipage}{0.5\hsize}
  \begin{center}
   \includegraphics[width=7.0cm]{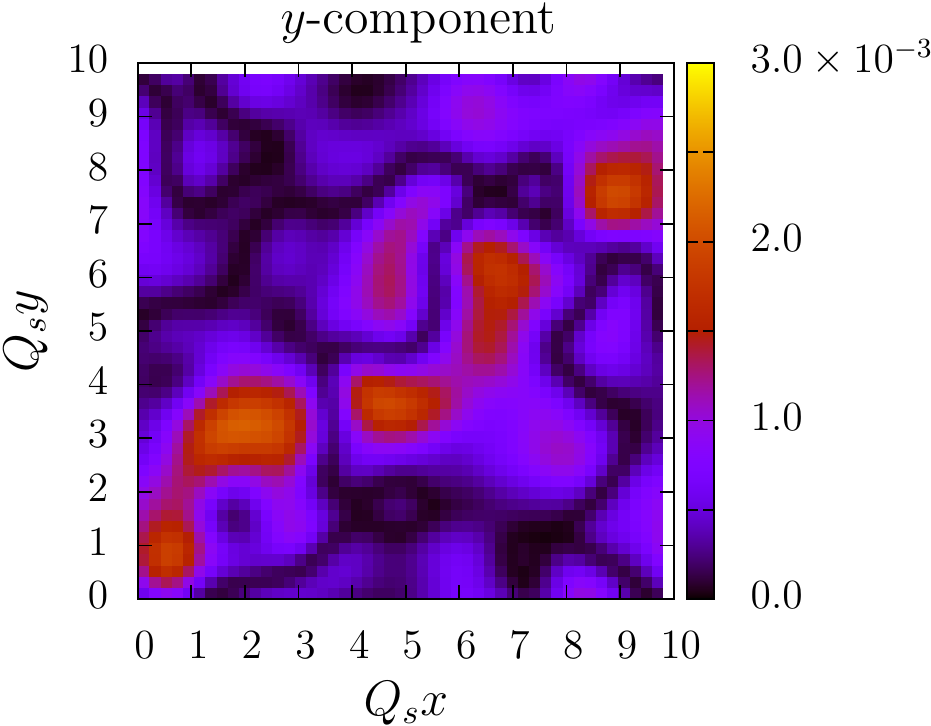} 
  \end{center}
 \end{minipage} 
 \end{tabular} \\[+10pt]
  \begin{center}
   \includegraphics[width=7.0cm]{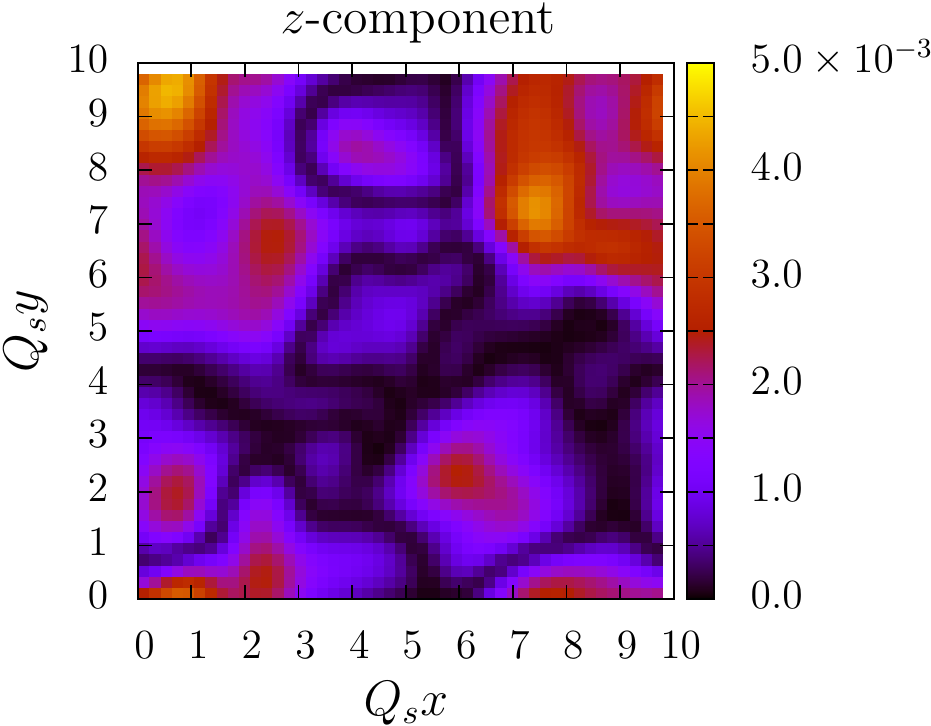} 
  \end{center}
 \caption{\label{fig:su3current} 
  The transverse coordinate dependence of the absolute value of the EM current divided by the EM coupling,
  $\left| J_\text{{\tiny EM}}^i \right| /(e Q_s^3)$, for SU(3)${}_c$ at the time $Q_s t=5$. 
  (Density plots with color bars in the right side of each figure.)
  Parameters are the same as those in Fig.~\ref{fig:emcsq}.  }
\end{figure*}

Under these classical gauge fields, we have solved the Dirac equation \eqref{Dirac2} and
computed the EM current \eqref{EMc1}. 
Because the color directions of the background gauge fields are random in the transverse plane, 
the space average of the EM current is nearly zero. 
In Fig.~\ref{fig:emcsq}, we plot the space average of the square of the EM current
for SU(2)${}_c$ (Left) and SU(3)${}_c$ (Right) as a function of time. 
Unlike the uniform field case, 
a nonzero current is induced even for SU(2)${}_c$. 
However, the current for SU(2)${}_c$ is several orders smaller than that for SU(3)${}_c$. 
In particular, the longitudinal component is smaller than the transverse ones in the case of SU(2)${}_c$
although the initial background electric field is along the longitudinal direction. 
While the argument of the cancellation of the EM currents between two colors is
strict only for a uniform system, 
this result indicates that there is still tendency of the cancellation even in the inhomogeneous system.
The nonzero but relatively small current originates in the violation of the Abelianization argument 
in nonuniform fields.
Hence, the transverse  components are larger than the longitudinal one for SU(2)${}_c$.  
In contrast, for SU(3)${}_c$, the cancellation is not perfect even in uniform fields, as shown in the previous section.
Therefore, there is no reason the EM current is suppressed like the SU(2)${}_c$ case. 
Indeed, for SU(3)${}_c$, the longitudinal current is dominant over the transverse ones. 
Figure \ref{fig:su3current} shows the transverse coordinate dependence of the absolute value of 
the EM current for SU(3)${}_c$ at the time $Q_s t=5$. 
Reflecting the inhomogeneity of the background gauge field, 
the current has inhomogeneity whose typical scale is the order of $1/Q_s$.

\begin{figure}[tbp]
\begin{center}
\includegraphics[clip,width=9.0cm]{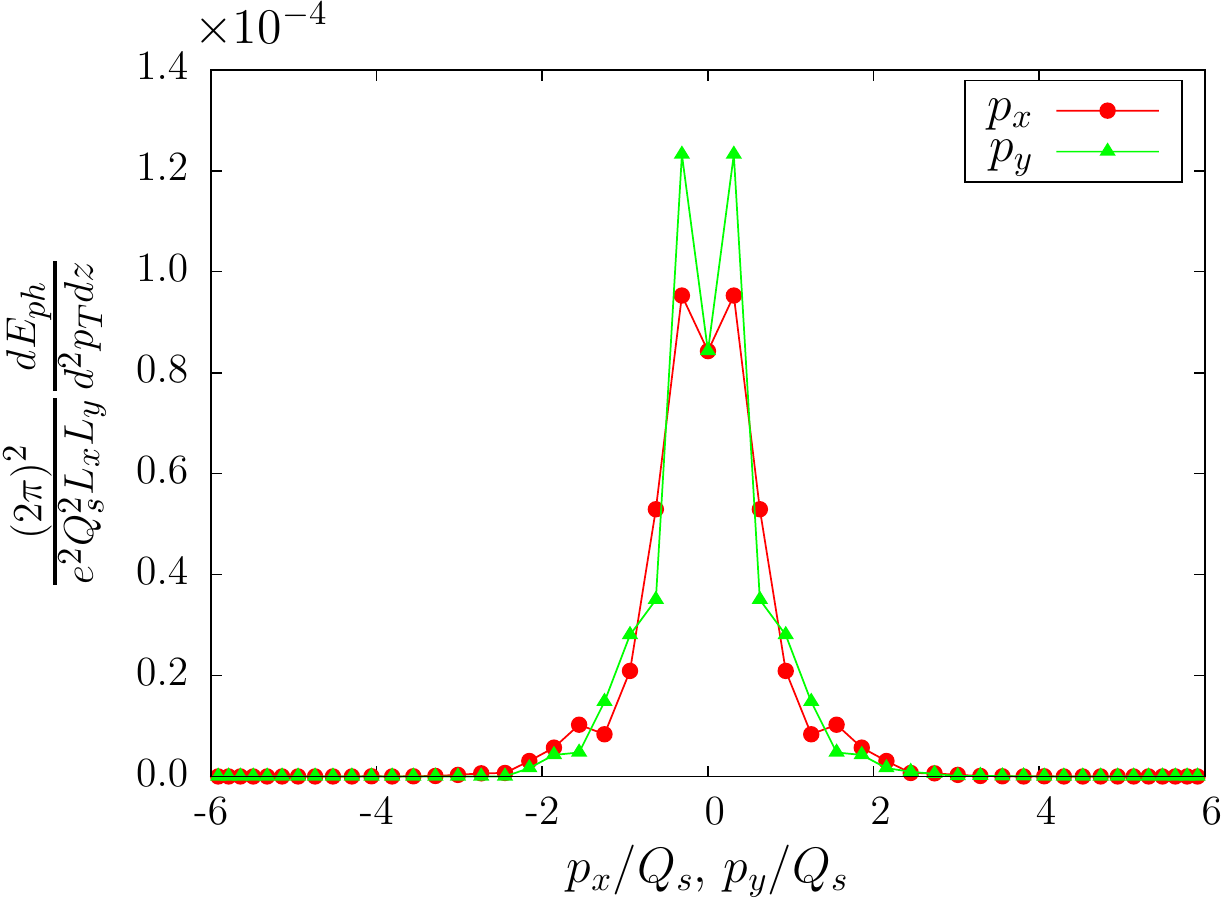}
\caption{The photon energy spectrum \eqref{ene_spec} for SU(3)${}_c$ at the time $Q_s t=5$. 
The $p_x$-distribution with fixed $p_y=0$ and the $p_y$-distribution with fixed $p_x=0$ are compared.
Parameters used for this computation are 
$m/Q_s =0.1$, $Q_s L_x =Q_s L_y =20$, $Q_s L_z =15$, $N_x =N_y =64$, and $N_z =48$. }
\label{fig:photonspec1}
\end{center}
\end{figure}

Once an EM current is induced, it can emit photons.
Photons emitted from
a coherent EM current (nonzero one-point expectation of the EM current operator)
form coherent classical EM fields.
This process is simply described by the classical Maxwell equation coupled to 
the EM current:
\begin{equation} \label{Maxwell}
\frac{d\mathbf{E}}{dt} -\nabla \times \mathbf{B} = -\mathbf{J}_\text{{\tiny EM}} \, . 
\end{equation}
We have numerically solved the lattice version of the Maxwell equation 
with the EM current induced by the SU(3)${}_c$ random fields. 
From the classical EM fields, we can compute the photon energy spectrum by the following equation:
\begin{equation} \label{ene_spec}
\frac{dE_{ph}}{d^2 p_T dz} = \frac{1}{2(2\pi)^2} \left[ \left| \tilde{\mathbf{E}} (\bp_T )\right|^2 
 +\left| \tilde{\mathbf{B}} (\bp_T )\right|^2 \right] \, ,
\end{equation}
with
\begin{gather}
\tilde{\mathbf{E}} (\bp_T ) = \int \! d^2 x_T \, \mathbf{E} (\bx_T ) e^{-i\bp_T \cdot \bx _T} \, , \\
\tilde{\mathbf{B}} (\bp_T ) = \int \! d^2 x_T \, \mathbf{B} (\bx_T ) e^{-i\bp _T \cdot \bx _T} \, .
\end{gather}
Because the system is uniform in the $z$-direction, we have not done the Fourier transform in $z$. 
If we do it, the $p_z$-distribution becomes the delta function $\delta (p_z)$.  
Figure \ref{fig:photonspec1} shows the photon energy spectrum \eqref{ene_spec} at the time $Q_s t=5$.
As the background color field is characterized by the scale $Q_s$, 
the photon energy spectrum has a width determined by the same scale $Q_s$. 
In the figure, $p_x$-distribution with fixed $p_y=0$ and $p_y$-distribution with fixed $p_x=0$ are compared. 
Reflecting the inhomogeneous structure of the background, 
the two distributions differ especially at the low momentum region $p_T \ltsim Q_s$.  

\begin{figure}[tbp]
\begin{center}
\includegraphics[clip,width=9.0cm]{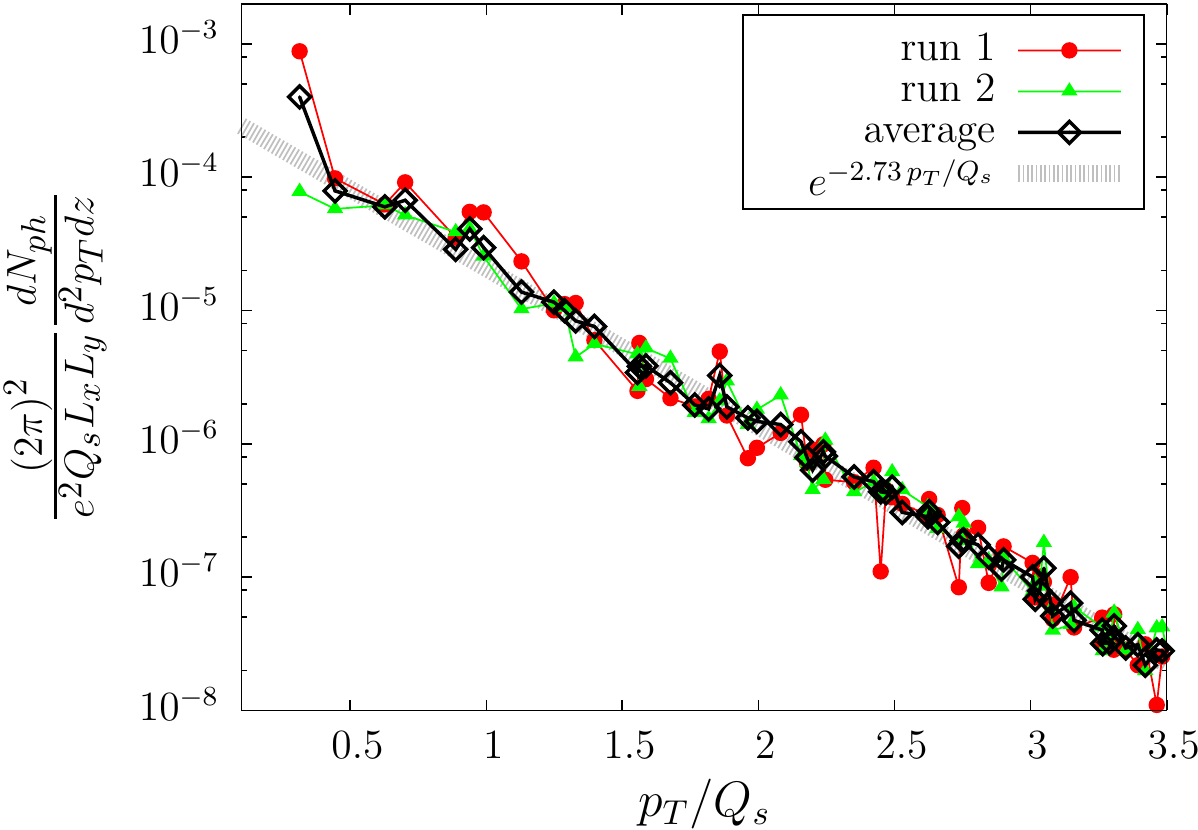}
\caption{The photon number spectrum \eqref{num_spec} as a function of $p_T$.
Results of two runs with different random initial backgrounds and
the average over four runs are plotted. 
Parameters used for these computations are the same as those in Fig.~\ref{fig:photonspec1}.  
A function $3.2\times 10^{-4} \,  e^{-2.73 p_T/Q_s}$ obtained by a fitting at the region $1< p_T/Q_s <4$
is plotted as well. }
\label{fig:photonspec2}
\end{center}
\end{figure}

In Fig.~\ref{fig:photonspec2}, the photon number spectrum defined by
\begin{equation} \label{num_spec}
\frac{dN_{ph}}{d^2 p_T dz} = \frac{1}{\omega} \frac{dE_{ph}}{d^2 p_T dz} \, ,
\end{equation}
with $\omega =p_T =\sqrt{p_x^2 +p_y^2}$, is plotted as a function of $p_T$. 
The $p_T$-distribution is computed by taking average over points $(p_x ,p_y)$
which give the same $p_T$. 
To show the dependence on the random background field configurations, 
the results of two runs with different initial color field configurations are plotted,
as well as the average over four runs.
At the low momentum region, $p_T < Q_s$, the spectrum fluctuates relatively largely run by run,
in accordance with the observation we made in Fig.~\ref{fig:photonspec1}. 
At the high momentum region, 
the dependence on the random filed configurations is relatively small, 
and the spectrum shows an exponential behavior. 
In Fig.~\ref{fig:photonspec2}, a function proportional to $e^{-2.73 p_T/Q_s}$,
which is obtained by a fitting at the region $1< p_T/Q_s <4$, is also plotted.  
It is worth noting that this exponential function resembles a thermal distribution $e^{-\omega/T}$
although the source of the photons is far away from thermal equilibrium.

\section{Summary and discussions} \label{sec:summary}
In this paper, we have investigated the induction of electromagnetic (EM) currents
associated with the nonperturbative quark production from classical color fields. 
In weak coupling and strong field regimes, which may be realized in the initial stage
of heavy-ion collisions,  
the quark production in the leading order with respect to the coupling $g$ 
is described by the retarded solutions (mode functions) of
the Dirac equation under the classical gauge field. 
Once the mode functions are obtained, 
the EM current can be computed from them. 
We have numerically solved the Dirac equation on real-time lattice. 

To clarify the significant difference between the SU(2)${}_c$ and the SU(3)${}_c$ theories 
for the induction of EM currents,  
we have considered, at first, uniform and constant color electric fields as background fields.
This field configuration enables us to simplify the color structure of the theory by 
diagonalization.
Thanks to this simplification, it can be easily shown that 
a net EM current is not generated in uniform SU(2) color fields
since the contributions from two colors cancel each other. 
In contrast, in uniform SU(3) color fields a nonzero net EM current is induced depending on
the color direction of the background electric field.

After confirming that our lattice numerical calculation in the uniform color field
reproduces known analytic solutions, 
we have applied our numerical method to more realistic inhomogeneous backgrounds. 
Motivated by the Glasma, 
we have introduced fluxtube-like color electric fields which are randomly distributed in the transverse plane
[Eq.~\eqref{randE}]. 
Under such inhomogeneous gauge fields, the argument of the exact cancellation of the EM current for SU(2)${}_c$
cannot be applied,
and thus nonzero EM current is generated even for the SU(2)${}_c$ case. 
However, the EM current for SU(2)${}_c$ is several orders smaller than that of SU(3)${}_c$,
because there is still a tendency of the cancellation between two colors.  
Therefore, 
it is important to pay attention to the difference between SU(2)${}_c$ and SU(3)${}_c$,
when one studies the induction of EM current and subsequent photon production. 
Using the EM current obtained by numerical computations, 
we further calculated the EM fields induced through the Maxwell equation. 
From the EM fields, the photon spectrum has been computed. 
Because the background color fields are characterized only by a single scale $Q_s$, 
the photon spectrum has a width given by the same scale $Q_s$. 
Reflecting the random nature of the background color fields, 
the photon spectrum fluctuates relatively largely at small momentum region, $p_T <Q_s$. 
We have found that the photon spectrum shows, at the region $p_T \gtsim Q_s$,
an exponential behavior $e^{-p_T/T_\text{eff}}$ with $T_\text{eff} \simeq Q_s/3$. 

\begin{figure*}[tbp]
 \begin{center}
 \begin{minipage}{0.3\hsize}
  \begin{center}
   \includegraphics[width=4.0cm]{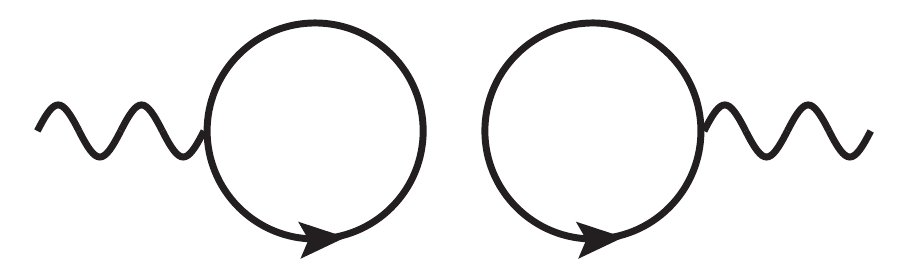} 
  \end{center}
 \end{minipage}
 \begin{minipage}{0.3\hsize}
  \begin{center}
   \includegraphics[width=4.0cm]{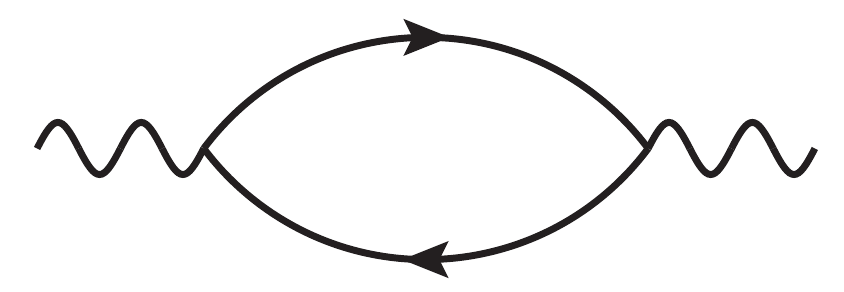} 
  \end{center}
 \end{minipage} 
 \end{center}
 \caption{\label{fig:diagrams} 
 The leading order diagrams for the photon 2-point function $\langle \hat{A}_\mu (x) \hat{A}_\nu (x)\rangle$. 
 Solid liens denote quark propagator dressed by background non-Abelian gauge fields. 
 Left: Disconnect diagram that represents the classical process of photon emission,
 which we have studied in this paper. 
 One loop tadpole corresponds to the one point expectation of the EM current operator,
 $\langle \hat{J}_\text{{\tiny EM}}^\mu (x)\rangle$. 
 Right: Connected diagram that describes a genuine quantum process of photon production,
 which is not considered in this paper. }
\end{figure*}

There are several directions of possible extensions of this study. 

While we have dealt with fixed  geometry, 
the systems of relativistic heavy-ion collisions show nearly boost-invariant expansion
in the longitudinal beam direction. 
The boost-invariantly expanding system can be conveniently described by  
the $\tau$-$\eta$ coordinates. 
Once we take account of the expanding geometry, 
we can utilize the McLerran--Venugopalan model \cite{McLerran:1993ni,McLerran:1993ka} 
for the initial condition of the classical background
fields instead of the random fields we introduced by hand. 
The initial condition for the quark mode functions which is compatible with 
the background fields provided by the color glass condensate framework
has been derived in Ref.~\cite{Gelis:2015eua}. 

It is also interesting and important to go beyond the leading order description in the coupling $g$. 
The next-to-leading order contributions for the quark production can be computed by 
adding fluctuations to the initial condition for the classical gauge fields
and considering the back reaction from produced quarks to the background color fields \cite{Kasper:2014uaa}. 
If the gauge field fluctuations are involved, the background gauge field is no longer 
uniform in any direction.
In this case, the numerical cost to solve the Dirac equation becomes unfavorably expensive,
and thus we may need to resort to the statistical method to sample the mode functions. 

In this paper, we have considered only one family of quark flavor. 
Let us comment on how our results are modified by the inclusion of multi-flavors
in the case of the uniform electric field.
In the leading order of the strong field and weak coupling approximation, 
quarks with different flavors are produced independently.
Therefore, we just need to sum up the EM currents contributed from each flavor. 
The quark masses shown in Fig.~\ref{fig:coefficient}, 
$m/\sqrt{gE_0} = 0.05, 0.65, 2.1$, correspond to strange, charm, bottom, respectively,
if we assume the field strength is related with the saturation scale as $gE_0 =Q_s^2$,
and take the value $Q_s =2$ GeV. 
Because the line of $m/\sqrt{gE_0} = 0.05$ is overlapped by that of $m/\sqrt{gE_0} = 0$,
strange and lighter flavors have the same yield. 
Since the sum of the EM charges of up ($+2e/3$), down ($-e/3$), and strange ($-e/3$) is zero, 
the total EM current induced by these three light flavors is vanishing. 
As a consequence, there remains only the contribution from charm quark. 
Since this argument is based on Eq.~\eqref{EMc_model} 
which is valid only for the uniform electric field, 
consideration of a realistic situation is needed to make a robust conclusion. 

The most important extension of this study would be the computation of 
a genuine quantum contribution for photon production. 
The photon spectrum can be related with the photon two point function
$\langle \hat{A}_\mu (x) \hat{A}_\nu (y) \rangle$. 
In the leading order with respect to the coupling $e$, 
there are two kinds of diagrams, shown in Fig.~\ref{fig:diagrams}.
In this paper, we have computed the classical EM fields by solving the Maxwell equation
which couples to the one-point expectation of the EM current operator,
and the photon spectra have been calculated from the classical EM fields. 
This classical contribution corresponds to the disconnected diagram shown in the left of Fig.~\ref{fig:diagrams}. 
The connected diagram shown in the right represents a genuine quantum process
in a sense that it is related with the connected two-point current correlation function
$\langle \hat{J}_\text{{\tiny EM}}^\mu (x) \hat{J}_\text{{\tiny EM}}^\nu (y) \rangle_\text{c}$,
and it is impossible to construct it from the classical current
$\langle \hat{J}_\text{{\tiny EM}}^\mu (x) \rangle$. 
Although the connected contribution is the same order in the coupling $e$ as the disconnected one,
it is not computed in this paper. 
It is important to investigate this genuine quantum contribution for the photon production, 
in addition to the classical contribution studied in this paper. 
In particular, 
the cancellation between different colors/flavors may not happen in this contribution,
because the two vertexes are connected in the right diagram. 
However, the computation of the genuine quantum contribution 
in the real-time non-equilibrium quantum field theory
involves several problems, 
which demand further investigation.
The present paper provides a basis for these future investigations.

\section*{Acknowledgements}
The author would like to thank J. Berges, F. Gelis, and N. M\"{u}ller for useful discussions 
related to this study. 
He also acknowledges the hospitality of the members of BNL Nuclear Theory Group and 
RIKEN BNL Research Center, 
where part of this study was conducted.

\appendix
\section{Numerical method} \label{sec:numerical}
\subsection{Lattice discretization} \label{subsec:lattice}
Except a few examples, 
analytic solutions of the Dirac equation under inhomogeneous gauge fields are hardly available. 
Therefore, one needs to resort to numerical methods. 
In order to numerically solve the classical Yang--Mills equation and the Dirac equation, 
we discretize the 3-dimensional space $V=L_x \times L_y \times L_z$ into 
$N_x \times N_y \times N_z$-lattice sites. 
The space coordinates are labeled by
\begin{equation}
\bx = (a_x n_x , a_y n_y ,a_z n_z ) \, ,
\end{equation}
where $n_i = 1,2, \cdots ,N_i$ and $a_i =L_i/N_i$ $(i=x,y,z)$. 
We assume the periodic boundary condition.
The plane wave is discretized as
\begin{equation}
e^{i\bp \cd \bx } = \exp \left\{ 2\pi i \left( \frac{k_x n_x}{N_x } +\frac{k_y n_y}{N_y } +\frac{k_z n_z}{N_z } \right) \right\} \, ,
\end{equation}
where $k_i =-\frac{N_i}{2} +1 ,\cdots ,-1,0,1,\cdots ,\frac{N_i }{2} $. 

We take the temporal gauge $A_0^a =0$ and keep the time as a continuum variable. 
In this case, the gauge fields are described by link variables $U_i (x) =\exp \left( iga_i A_i (x) \right)$
and electric fields $E_i (x)$.
The equations of motion for these variables are
\begin{equation} \label{eq_U}
\partial_t U_i (x) = iga_i E_i (x) U_i (x) \, , 
\end{equation}
and
\begin{align}
\partial_t E_i (x) 
= -\frac{1}{g} \frac{1}{a_i} \underset{(j\neq i)}{\sum_{j=1}^3} \frac{1}{a_j^2} 
 \left\{ \text{Im} \left[ U_{i,j} (x) +U_{i,-j} (x) \right] -\text{(trace)} \right\} \, , \label{eq_E}
\end{align}
where $U_{i,j} (x)$ and $U_{i,-j} (x)$ are the plaquettes defined by
\begin{equation}
U_{i,j} (x) = U_i (x) U_j (x+\hat{i}) U_i^\dagger (x+\hat{j}) U_j^\dagger (x) \, , 
\end{equation}
\begin{equation}
U_{i,-j} (x) = U_i (x) U_j^\dagger (x+\hat{i}-\hat{j}) U_i^\dagger (x-\hat{j}) U_j (x-\hat{j} ) \, ,
\end{equation}
and \lq\lq $-$(trace)\rq\rq\ means
\begin{equation}
X-\text{(trace)} = X-\frac{1}{N_c} \text{tr} X \cdot 1 \hspace{20pt} (\text{for } X\in \text{SU($N_c$)}) .
\end{equation}
The lattice Yang--Mills equations \eqref{eq_U} and \eqref{eq_E} preserve the Gauss's law
\begin{equation} \label{Gauss}
\sum_{i=1}^3 \frac{1}{a_i} \left[ E_i (x) -U_i^\dagger (x-\hat{i} ) E_i (x-\hat{i} ) U_i (x-\hat{i} ) \right] = 0 \, . 
\end{equation}

Since derivative appears linearly in the Dirac equation, the derivative for the Dirac field must be replaced
by the center difference to maintain the hermiticity of the theory: 
\begin{equation}
D_i \psi (x) = \frac{1}{2a_i } \left[ U_i (x) \psi (x+\hat{i} ) -U_i^\dagger (x-\hat{i} ) \psi (x-\hat{i} ) \right] \, .
\end{equation}
Using this covariant difference, we need to solve the Dirac equation for the mode functions, 
\begin{equation} \label{Dirac2}
\left[ i\gamma^0 \partial_0 +i\gamma^i D_i -m \right] \psi_{\bp ,s,c}^\pm (x) = 0  \, , 
\end{equation} 
as well as the lattice Yang--Mills equations \eqref{eq_U} and \eqref{eq_E}. 
We assume the null initial condition for the gauge field: $U_i (t_0)=1$. 
Hence the quark mode functions must obey the free initial condition.

On the lattice, the vacuum expectation value of the EM current operator 
$\hat{J}^i (x) =e\overline{\hat{\psi}} (x) \gamma^i \hat{\psi} (x)$
can be expressed in terms of the mode functions as 
\begin{equation} \label{EMc1}
J_\text{{\tiny EM}}^i 
 = e\sum_{s,c} \frac{1}{V} \sum_{\bp} \text{Re} \left[ 
   \overline{\psi}_{\bp ,s,c}\hspace{-17pt}{}^- \hspace{10pt}(x) \gamma^i U_i (x) \psi_{\bp ,s,c}^- (x+\hat{i}) \right] .
\end{equation}
Note that only the negative energy mode functions are necessary to compute such expectations.

The numerical cost to compute the mode functions $\psi_{\bp ,s,c}^- (x)$ is proportional to
$(N_x N_y N_z )^2$ because the mode functions depend on both $\bx$ and $\bp$. 
To reduce this rather expensive numerical cost, one can employ a Monte Carlo method 
\cite{Borsanyi:2008eu,Gelis:2015eua}.
However, in this paper, we use the deterministic method because we treat gauge fields which are uniform in 
one direction at least, of which nature reduces the numerical cost.
Under a gauge field uniform in the $z$-direction, the $z$-dependence of 
the mode functions $\psi_{\bp ,s,c}^- (x)$ is always $e^{-ip_z z}$. 
Therefore, the $z$-dependence need not be computed numerically. 
The scaling with the lattice size for the numerical cost to compute the mode functions is modified to $(N_x N_y )^2 N_z$. 
A merit of the deterministic method is that it is free from statistical errors. 
This is beneficial especially when one computes local expectations under inhomogeneous backgrounds.  

\subsection{Treatment of doublers} \label{subsec:doubler}

\begin{figure}[tbp]
\begin{center}
\includegraphics[clip,width=8.0cm]{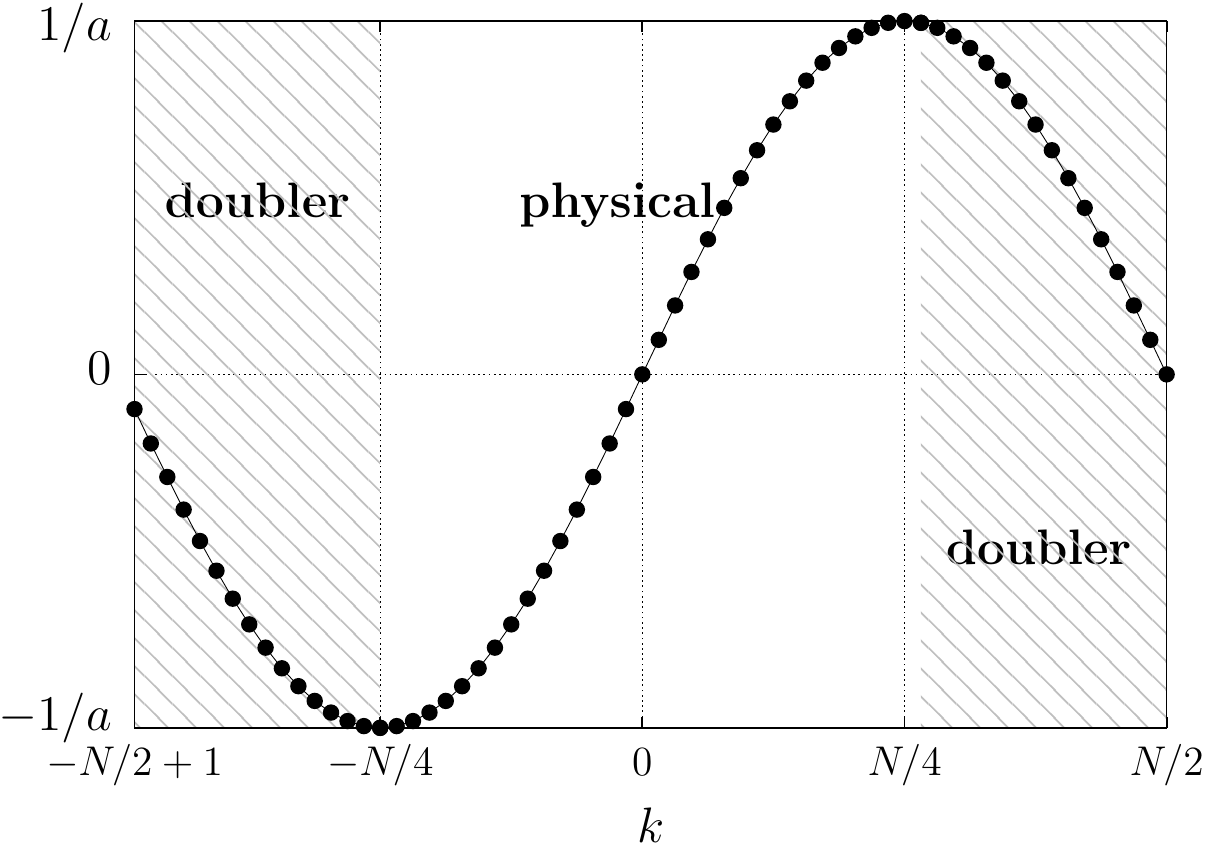}
\caption{A schematic plot of the lattice momentum \eqref{latmom} as a function of the integer $k$. }
\label{fig:latmom}
\end{center}
\end{figure}

As is well known, naively discretized fermionic fields suffer from the problem of doublers. 
The lattice momentum associated with the center difference 
\begin{equation} \label{latmom}
\hat{p}_i = \frac{1}{a_i } \sin \left( 2\pi \frac{k_i }{N_i } \right) 
\end{equation}
has 2-fold degeneracy. 
It has two zeros at $k_i =0$ and $k_i =N_i /2$ as Fig.~\ref{fig:latmom} shows. 
In 3-dimensional momentum space, there is 8-fold degeneracy in total. 
We call the region
\begin{equation}
\Lambda_\text{phys} = \left\{ (k_x ,k_y ,k_z ) \bigg| k_i =-\frac{N_i}{4} +1 ,\cdots ,0, \cdots ,\frac{N_i}{4} \right\}
\end{equation}
the physical region, and the remain the doubler region $\Lambda_\text{doubler}$. 
A way to decouple the doubler modes from the dynamics is adding the Wilson term to the Dirac equation:
\begin{align}
\left( i\gamma^0 \partial_0 +i\gamma^i D_i -m \right) \psi (x) 
 +\frac{r}{2} \sum_{i=1}^3 \frac{1}{a_i} 
 \left[ U_i (x) \psi (x+\hat{i} ) -2\psi(x) +U_i^\dagger (x-\hat{i}) \psi (x-\hat{i} ) \right] 
 = 0 \, . 
\end{align}
However, we do not employ the Wilson term by the following reason. 

Because the Wilson term contains the gauge link, it contributes to the EM current. 
In addition to the normal contribution \eqref{EMc1}, we have
\begin{equation} \label{EMcW}
J_\text{{\tiny EM,W}}^i 
 = er\sum_{s,c} \frac{1}{V} \sum_{\bp} \text{Im} \left[ 
   \overline{\psi}_{\bp ,s,c}\hspace{-17pt}{}^- \hspace{10pt}(x) U_i (x) \psi_{\bp ,s,c}^- (x+\hat{i}) \right] .
\end{equation}
In the continuum limit, the contribution from the Wilson term vanishes since $\overline{\psi}(x) \psi(x)$ is real. 
However, with a nonzero lattice spacing and under a nontrivial gauge field configuration, 
this contribution stays nonzero and contaminates the normal contribution. 
In particular, for the EM current, the normal contribution may be small or zero due to the cancellation
between different colors as discussed in Sec.~\ref{sec:uniform}, 
and thus the Wilson term contribution can be the same order to the normal contribution. 
To circumvent this problem, we employ an alternative way to decouple the doubler modes.

Because we directly handle the mode functions with the specific initial conditions, 
we can restrict the momentum $\bp$ 
of the mode function $\psi_{\bp ,s,c}^- (x)$ to the physical modes, $\bp \in \Lambda_\text{phys}$. 
By this, the mode functions which have the momentum index $\bp \in \Lambda_\text{doubler}$ can be simply excluded from the dynamics.
However, the effects of the doubler modes may not be completely eliminated by this procedure 
since the mode functions have also $\bx$-dependence. 
If we Fourier-transform the negative-energy mode functions as
\begin{equation}
\tilde{\psi}_{\bp ,s,c}^- (t, \bp^\prime ) = \frac{1}{N^3} \sum_{\bx} e^{i\bp^\prime \cd \bx} \psi_{\bp ,s,c}^- (x) \, , 
\end{equation}
then the momentum $\bp^\prime$ spans over the full momentum space $\Lambda_\text{phys} \oplus \Lambda_\text{doubler}$. 
There is still a possibility that the doubler modes contaminate the mode functions 
through the $\bx$-dependence. 
At the initial time, the mode functions are proportional to the plane wave $e^{-i\bp \cd \bx}$,
so that $\tilde{\psi}$ is diagonal in the momentum space:
\begin{equation}
\tilde{\psi}_{\bp ,s,c}^- (t=0, \bp^\prime ) \propto \delta_{\bp ,\bp^\prime } \, . 
\end{equation}
Because we have restricted the momentum $\bp$ to the physical modes, 
also the momentum $\bp^\prime$ is in the physical region. 
In this sense, the effects of doublers are completely eliminated at the initial time 
just by the limitation $\bp \in \Lambda_\text{phys}$. 
By the time evolution, the contamination by the doubler modes may come in. 
This point becomes apparent if we Fourier-transform the Dirac equation:
\begin{equation}
\begin{split}
&\left( i\gamma^0 \partial_0 -m \right) \tilde{\psi}_{\bp ,s,c}^- (t, \bp^\prime ) \\
&\hspace{20pt} 
  +i\sum_{j=1}^3 \gamma^j \frac{1}{2a_j} \sum_\bq 
  \left[ e^{-i(p^{\prime j} -q^j )a_j } \tilde{U}_j (\bq ) 
  -e^{ip^{\prime j} a_j } \tilde{U}_j^\dagger (-\bq ) \right] \tilde{\psi}_{\bp ,s,c}^- (t, \bp^\prime -\bq )
 = 0 \, , 
\end{split}
\end{equation}
where $\tilde{U}_i (\bq )$ is the Fourier transform of the link variable. 
If the gauge field is inhomogeneous, $\tilde{U}_i (\bq )$ can carry nonzero momentum
and the gauge field may kick out quarks from the physical momentum region to the doubler region. 
By explicitly computing $\tilde{\psi}_{\bp ,s,c}^- (t, \bp^\prime )$, we can monitor
how the contamination occurs. 

\begin{figure*}[tbp]
\begin{tabular}{cc}
 \begin{minipage}{0.5\hsize}
  \begin{center}
   \includegraphics[width=7.0cm]{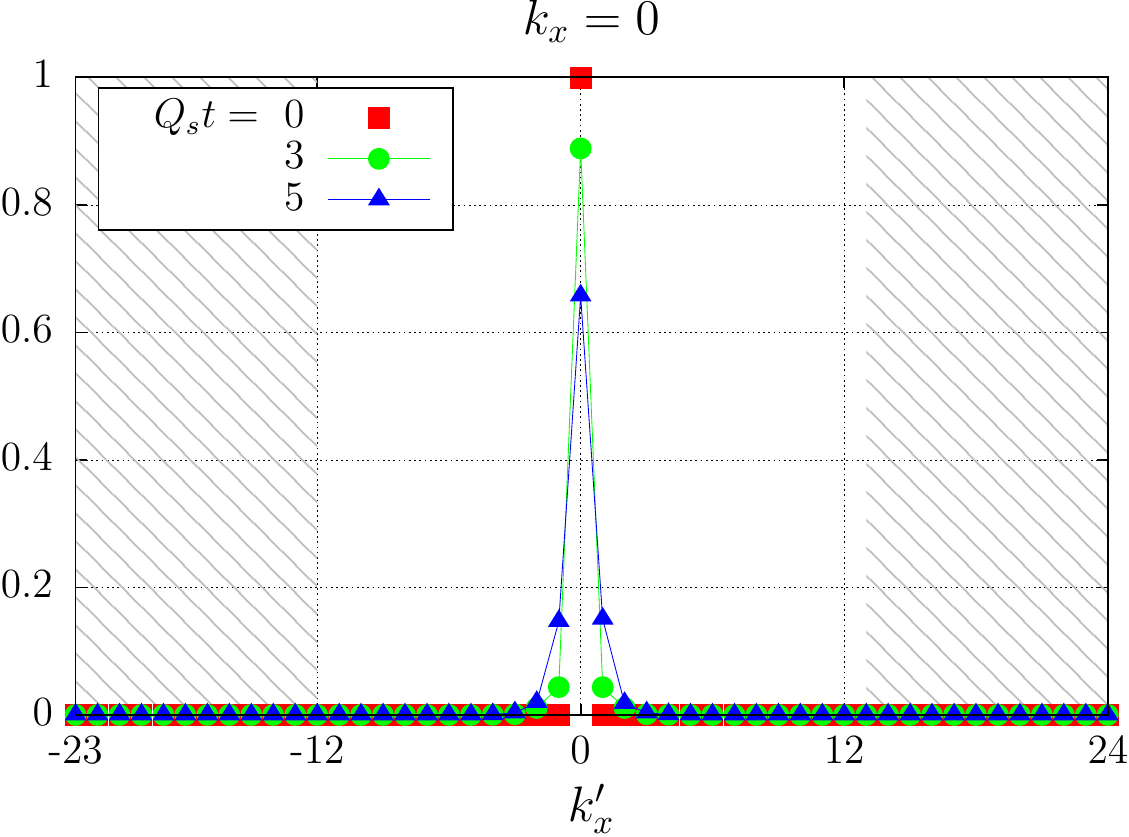}
  \end{center}
 \end{minipage}
 \begin{minipage}{0.5\hsize}
  \begin{center}
   \includegraphics[width=7.0cm]{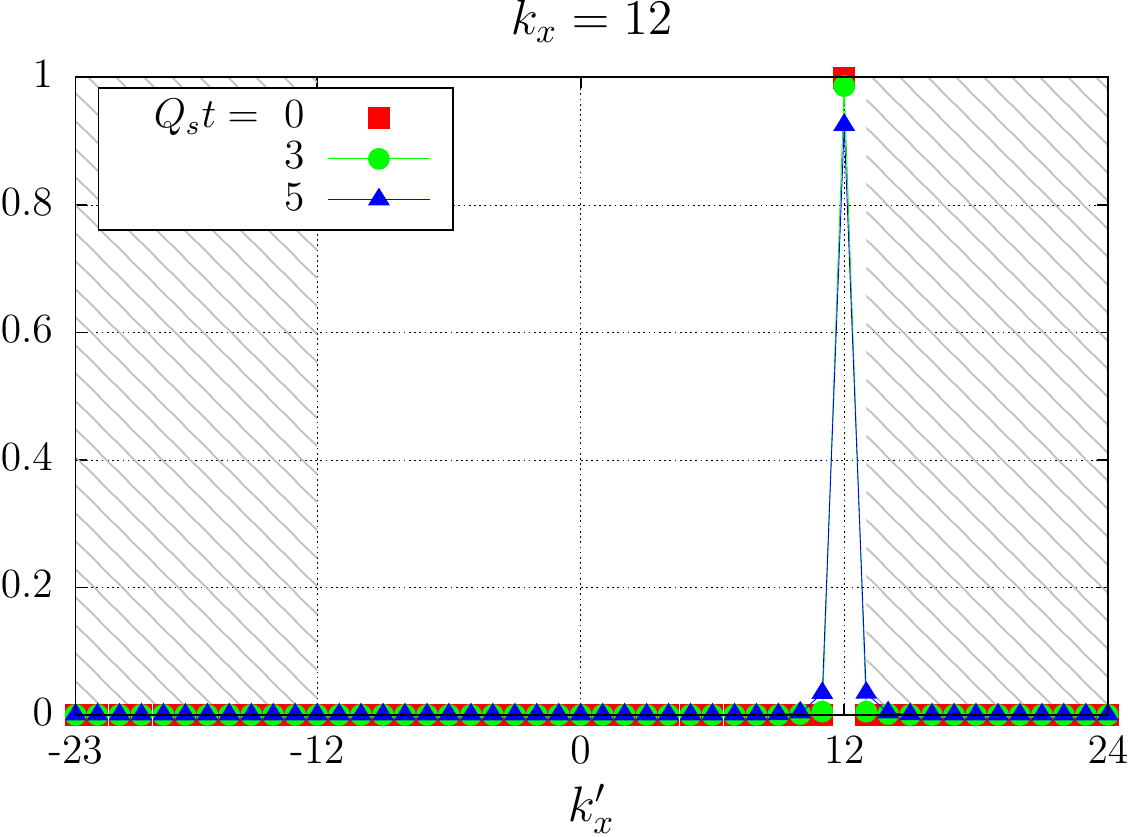} 
  \end{center}
 \end{minipage} 
\end{tabular}
 \caption{\label{fig:dbl} Plots of $\sum_{k_y^\prime} M(t; k_x ,k_y=0 ;k_x^\prime ,k_y^\prime )$
 for (a) $k_x=0$ and (b) $k_x=12$.
 This quantity equals to $\delta_{k_x ,k_x^\prime}$ at the initial time $Q_s t=0$. 
 Shaded regions represent the doubler region. 
 Parameters used for this computation are $m/Q_s=0.1$, $Q_s L_x =Q_s L_y =10$, $Q_s L_z =15$,
 and $N_x =N_y =N_z =48$.}
\end{figure*}

For that purpose, we consider the following quantity:
\begin{equation}
M(t; k_x ,k_y ;k_x^\prime ,k_y^\prime ) 
 = \frac{1}{N_z} \sum_{k_z \in \Lambda_\text{phys}} \sum_{k_z^\prime} \frac{1}{2N_\text{c}} \sum_{s,c}  
   \tilde{\psi}_{\bp ,s,c}^{-\, \dagger} (t, \bp^\prime ) \tilde{\psi}_{\bp ,s,c}^- (t, \bp^\prime ) ,
\end{equation}
where $k_x$ and $k_y$ ($k_x^\prime$ and $k_y^\prime$) are integers corresponding to the momenta $p_x$ and $p_y$
($p_x^\prime$ and $p_y^\prime$), respectively. 
Because $\tilde{\psi}_{\bp ,s,c}^- (t, \bp^\prime )$ is always diagonal with respect to $p_z$ and $p_z^\prime$
under longitudinally uniform gauge fields we consider in this paper, 
the sum over the longitudinal momenta is already taken. 
Thanks to the normalization condition for the mode functions,  
the function $M$ satisfies
\begin{equation}
\sum_{k_x^\prime ,k_y^\prime} M(t; k_x ,k_y ;k_x^\prime ,k_y^\prime ) = 1 \, . 
\end{equation}
At the initial time, the function $M$ is totally diagonal,
\begin{equation} \label{iniM}
M(0; k_x ,k_y ;k_x^\prime ,k_y^\prime ) = \delta_{k_x ,k_x^\prime} \delta_{k_y ,k_y^\prime} \, . 
\end{equation}
It therefore vanishes if $(k_x^\prime ,k_y^\prime ) \in \Lambda_\text{doubler}$ 
since $(k_x ,k_y ) \in \Lambda_\text{phys}$. 
Figure \ref{fig:dbl} exhibits how the function $M$ evolves in time under the inhomogeneous gauge field
introduced in Sec.~\ref{sec:glasma}. 
Both of the strength and the extent of the inhomogeneity of this gauge field is characterized 
by the scale $Q_s$.    
As examples, $\sum_{k_y^\prime} M(t; k_x ,k_y;k_x^\prime ,k_y^\prime )$ with $(k_x ,k_y)=(0,0)$ and $(12,0)$
are shown. 
At the initial time $Q_s t=0$, the $k_x^\prime$-distribution is localized at $k_x^\prime =k_x$. 
Since the inhomogeneous background gauge field carries nonzero transverse momentum, 
the $k_x^\prime$-distribution gets broadened by time evolution. 
However, the extent of the broadening is an order of $Q_s$, 
which is a typical momentum the background field carries,
and far smaller than the cutoff momentum.
In particular, when $k_x^\prime$ is near to the border with the doubler region 
(the case of $k_x^\prime =12$, plotted in the right), 
the broadening is weaker than the case $k_x^\prime$ is at a low momentum region. 
As long as we restrict $(k_x,k_y)$ to the physical region, 
$(k_x^\prime ,k_y^\prime )$ at the doubler region is not significantly excited. 
In particular, the zero momentum point of the doubler ($k_x^\prime =24$ in this case)
is not at all activated. 
Although we have shown only two examples of the parameters $(k_x ,k_y)$, 
the broadening is weak for any value of $(k_x ,k_y)$, at least within the timescale of the order
of $1/Q_s$, which we are interested in. 
To conclude, by restricting the momentum $\bp$ in the index of the mode functions $\psi_{\bp ,s,c}^- (x)$
to the physical region, we can suppress the contamination by doubler modes%
\footnote{If one is interested in chiral anomaly, this method cannot be applied,
because this method amounts to having a cutoff which breaks gauge invariance.}. 
This would be not true if the background field is highly inhomogeneous, 
or if we continue the time evolution till a late time.
For example, if one reaches thermal equilibrium after a long time evolution, 
the physical modes and the doubler modes would be equally excited unless one uses other method to suppress doublers. 
However, in the background field and time scale we are interested in, the contamination by the doublers does not matter.

This method has an advantage in numerical cost. 
Because we restrict the momentum $\bp$ to the physical region,
which has the volume of $1/2^3$ of the whole momentum space, 
the numerical cost is reduced by factor eight.


\end{document}